\documentclass{aa}
\usepackage[varg]{txfonts}
\usepackage{rotating}
\usepackage{colortbl}
\usepackage{caption}
\usepackage{subcaption}

\begin{document}

\title{The shape of convective core overshooting from gravity-mode period spacings}

\author{M. G. Pedersen\inst{1}
  \and C. Aerts\inst{1,2}
  \and P. I. P{\'a}pics\inst{1}
  \and T. M. Rogers\inst{3,4}}

\institute{Instituut voor Sterrenkunde, KU Leuven, Celestijnenlaan 200D, 3001, Leuven, Belgium, \email{maygade.pedersen@kuleuven.be}
  \and Department of Astrophysics/IMAPP, Radboud University Nijmegen, 6500 GL, Nijmegen, The Netherlands
  \and Department of Mathematics \& Statistics, Newcastle University, UK
  \and Planetary Science Institute, Tucson, AZ, 85721, USA}

\date{Received Date Month Year / Accepted Date Month Year}

\abstract {The evolution of stars born with a convective core is highly dependent on the efficiency and extent of near core mixing processes, which effectively increases both the core mass and main-sequence lifetime. These mixing processes remain poorly constrained and therefore result in large uncertainties in the stellar structure and evolution models of such stars.} {We investigate to what extent gravity-mode period spacings in slowly pulsating B--type stars observed by the \emph{Kepler} mission can be used to constrain both the shape and extent of convective core overshoot and additional mixing in the radiative envelope.} {We compute grids of 1D stellar structure and evolution models for two different shapes of convective core overshooting and three shapes of radiative envelope mixing. The models in these grids are compared to a set of benchmark models to evaluate their capability of mimicking the dipole prograde g-modes of the benchmark models.} {Through our model comparisons we find that at a central hydrogen content of $X_\text{c} = 0.5$, dipole prograde g-modes in the period range 0.8-3 d are capable of differentiating between step and exponential diffusive overshooting. This ability disappears towards the terminal age main-sequence at $X_\text{c} = 0.1$. Furthermore, the g-modes behave the same for the three different shapes of radiative envelope mixing considered. However, a constant envelope mixing requires a diffusion coefficient near the convective core five times higher than chemical mixing from internal gravity waves to obtain a surface nitrogen excess of $\sim 0.5$ dex within the main-sequence lifetime.} {Within estimated frequency errors of the \emph{Kepler} mission, the ability of g-modes to distinguish between step and exponential diffusive overshooting depends on the evolutionary stage. Combining information from the average period spacing and observed surface abundances, notably nitrogen, could potentially be used to constrain the shape of mixing in the radiative envelope of massive stars.}

\maketitle 


\section{Introduction}\label{Sec:Introduction}

A wide range of fields in astronomy rely on the accurate predictions of stellar structure and evolution models. To name a few examples, the study of stars that end their lives in supernovae explosions depends on  knowledge of their progenitors, while intermediate-mass stars are particularly important for the age determination of stellar clusters. In stars with convective cores, the size of the core is effectively enlarged through mechanisms such as convective core overshooting. Such overshooting brings additional hydrogen into the core of the star and therefore directly impacts the final He core mass and main sequence (MS) lifetime as well as the evolution of the stars after the MS. Different shapes of the convective core overshooting are either more or less effective at bringing such additional material into the stellar core, and result in different He core masses at the end of the MS. Therefore, constraining both the extent and shape of convective core overshooting remains of prime importance for the evolution of stars with convective cores.

Isochrone fitting of stellar clusters in the Milky Way and the Magellanic Clouds has been used to constrain both the extent and the necessity of convective core overshooting to explain the position and morphology of the MS turnoff in Colour Magnitude Diagrams \citep[e.g.][]{Maeder1981,Aparicio1990,Meynet1993,Kozhurina-Platais1997,VandenBerg2004,Rosenfield2017}. Simultaneously fitting evolutionary tracks of binary stars have lead to estimates of not only convective core overshooting \citep[e.g.][]{Guinan2000,Groenewegen2007,Lacy2012,Prada2012,Stancliffe2015} but also its dependence on stellar mass \citep[e.g.][]{Ribas2000,Claret2016,Claret2017}. While both methods have provided rough constraints on the extent of the overshooting, its shape has not been calibrated so far.

Another approach to determine convective core overshoot is through asteroseismology. Constraints have been obtained from the ratios of small separations of radial and dipole modes to the large separation for solar-like oscillators on the MS \citep[e.g.][]{SilvaAguirre2011,Liu2014,Yang2015,Deheuvels2016}, the period spacings of mixed modes in RGB- \citep{Deheuvels2011,Montalban2013,Arentoft2017}, retired A--type \citep{Hjorringgaard2017} and $\delta$ Sct stars \citep{Lenz2010}, rate of period change of Cepheids \citep{Fadeyev2015} and seismic modelling of $\beta$ Cep stars \citep{Aerts2003,Pamyatnykh2004,Walczak2015}. However, just like the case of isochrone fitting and binary stars, none of these asteroseismics modelling efforts have been able to constrain the shape of convective core overshooting.

Asteroseismology of gravity-modes (g-modes) might hold the key to constrain not only the extent but also the shape of convective core overshooting. These pulsation modes, which are found in both $\gamma$ Doradus ($\gamma$ Dor) and Slowly Pulsation B--type (SPB) stars  on the MS \citep[see e.g.][]{Aerts2010}, probe the near core regions of the stars and are highly sensitive to the presence of chemical gradients in the interiors \citep{Miglio2008}. Such a gradient is developed naturally as stars more massive than $\sim 1.5 \ \text{M}_\odot$ evolve along the MS. During this evolution the convective core shrinks, leaving behind a chemical gradient \citep[e.g., ][Fig. 3.5 in Chapter 3]{Aerts2010} which causes spikes to appear in the Brunt-V\"{a}is\"{a}l\"{a} frequency and thereby leads to mode trapping \citep{Miglio2008}. Such mode trapping is seen directly as dips in the period spacing series of g-modes \citep[see e.g. Fig. 2 in][]{Miglio2008}, which are series displaying the period difference between modes of consecutive radial orders $n$ and the same spherical degree $l$ and azimuthal order $m$. Additional mixing processes near the convective core change the chemical gradient and thereby directly affect the resulting period spacing series. In the same way a change in the shape of the convective core overshooting results in a different shape of the chemical gradient and thereby makes it possible to use g-modes to constrain the shape of convective core overshooting.

\citet{Degroote2010} signified the first detection of a period spacing series in an SPB pulsator observed by the CoRoT space telescope \citep{Auvergne2009}. While deviations from a uniform period spacing were detected for this star, the seismic modelling only allowed for a lower limit estimate of the extent of the overshooting. Seismic modelling of independent g-modes detected in two SPBe stars likewise observed by CoRoT provided constraints on the extent of the overshooting although no period spacing series were detected \citep{Neiner2012}.

The four years of continuous photometric data provided by the \emph{Kepler} space mission \citep{Borucki2010}, resulting in a ten fold increase in frequency precision compared to CoRoT, allowed for the first detection
\citep{Papics2014,Papics2015} and detailed seismic modelling \citep{Moravveji2015,Moravveji2016ApJ...823..130M} of a period spacing series for two single SPB stars. In both cases an exponential description of the overshooting was favoured over a simple extension of the convective core, and additional mixing in the radiative envelope was required. An additional five SPB stars with period spacing series have recently been added to this sample \citep{Papics2017} and have yet to undergo similar detailed seismic modelling. In comparison, a total of 67 $\gamma$ Dor stars with period spacing series were detected in the sample of \emph{Kepler} stars by \citet{VanReeth2015}. These stars are likewise waiting to be modelled seismically, whereas a system of two hybrid $\delta$ Sct/$\gamma$ Dor binary pulsators observed by \emph{Kepler} was unable to provide constraints of the shape of the convective core overshooting from seismic modelling efforts of their period spacing series \citep{Schmid2016}. Nevertheless, due to the high frequency precision obtained for \emph{Kepler} stars we are now at a stage where g-modes hold the potential to distinguish between different shapes of near core mixing processes.

In this paper, the science questions we intend to answer are to what extent we can use g-mode pulsations in stars observed by the \emph{Kepler} telescope to distinguish between \textbf{a)} a step overshoot formulation versus exponential decaying diffusive mixing, and \textbf{b)} different shapes and efficiencies of extra diffusive chemical mixing in the radiative envelope, including information on expected surface nitrogen abundances. To carry out this investigation, grids of 1D stellar models are computed using the state-of-the-art 1D stellar structure and evolution code MESA \citep{Paxton2011,Paxton2013,Paxton2015} version r8118. For these grids, different shapes of the convective core overshooting and radiative envelope mixing (described in Sect. \ref{sec:MESA3OV}) are used. The pulsation mode properties of the stellar models are determined using the stellar oscillation code GYRE \citep{Townsend2013}, and compared to chosen benchmark models. Section \ref{sec:GeneralSetup} describes the general setup for the grids of stellar models as well as how the model comparison is carried out. In Sect. \ref{Sec:ShapeComparisons} we test the predictability of g-modes on differentiating between different shapes of overshooting and radiative envelope mixing. The combined probing capability of g-modes and surface nitrogen abundances is discussed in Sect. \ref{Sec:gmodesAndN14}. Finally, we summarize our conclusions in Sect. \ref{Sec:Conclusions}. These conclusions will form the basis of future asteroseismic modelling of \emph{Kepler} SPB stars. 

\section{Overshooting descriptions available in MESA}\label{sec:MESA3OV}

In all cases mentioned here, the temperature gradient in the overshoot region is the radiative one, $\nabla = \nabla_\text{rad}$. This implies that the thermal structure remains unchanged by the overshooting, and only the chemical mixing is affected as discussed in detail by \citet{Viallet2015}. The different shapes of overshooting described below are therefore only differentiating in the \textit{efficiency} of the chemical mixing that they introduce in the overshooting region. 

\subsection{Step overshooting}\label{Sec:StepOv}

The step overshoot is the simplest out of the three convective core overshoot descriptions available in MESA. It signifies the simplest way to go from a rigid convective boundary, resulting from the simplified 1D description of convection in the mixing length theory \citep{BohnVitense1958}, to allow the convective flow to penetrate into the radiative zones due to the inertia of the convective elements at the boundary. When instead a $\nabla = \nabla_\text{ad}$ is used in the overshooting region, this type of overshooting is also referred to as convective penetration \citep[see, e.g.,][ for further discussion on this]{Viallet2015}. Here we have used $\nabla = \nabla_\text{rad}$ in the overshooting region as mentioned above.

The step overshooting formalism implemented in MESA assumes that overshooting extends over a distance $\alpha_{ov} \cdot H_{p,cc}$ from the convective core boundary, $r_{cc}$, into the radiative envelope. $H_{p,cc}$ is the pressure scale height at $r_{cc}$ and $\alpha_{ov}$ is the extent of the overshooting, i.e. the parameter determining the size of the overshooting region. Here $r_{cc}$ is defined as the position at which $\nabla_\text{ad} = \nabla_\text{rad}$, and the mixing is assumed to be constant and instantaneous with the diffusion coefficient given by 

\begin{equation}
D_{OV} = D_0.
\label{eq:Dstep}
\end{equation}

\begin{figure*}
\centering
\includegraphics[width=\linewidth]{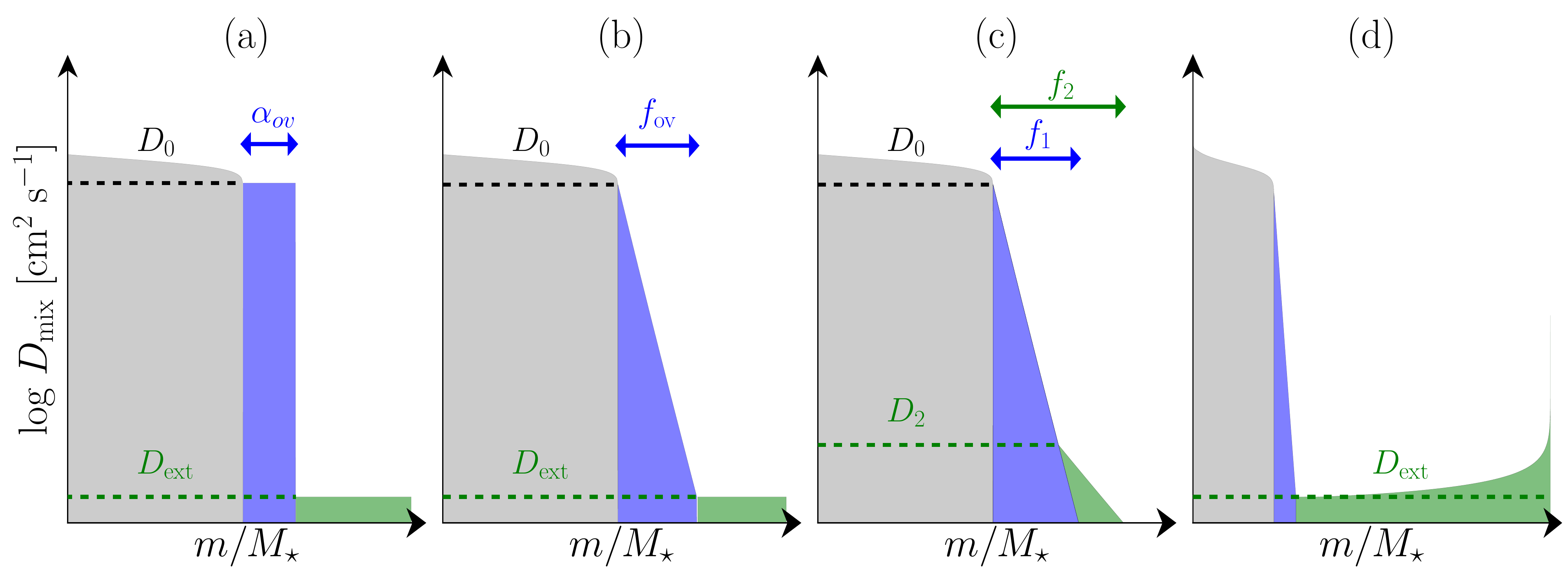}
\caption{Different shapes of internal mixing profiles. \textit{Grey} marks the convective core, \textit{blue} the overshooting region and \textit{green} the extra diffusive mixing in the radiative envelope. Panel (a) to (c)  has been zoomed in on the near core region while panel (d) shows the mixing profile from center to the surface of the star. In both panel (a) and (b) the extra diffusive mixing in the radiative envelope has been set constant. \textbf{Panel (a):} step overshoot. \textbf{Panel (b):} exponential overshooting. \textbf{Panel (c):} Extended exponential overshoot where the extension replaces the constant diffusive envelope mixing in panels (a) and (b). \textbf{Panel (d):} Exponential overshoot coupled to an extra diffusing mixing profile $D_\text{ext}(r)$ from \citet{Rogers2017} instead of a constant mixing (green dashed line).}
\label{fig:DmixDiagram}
\end{figure*}

\noindent Because the diffusive mixing coefficient goes to zero at $r_{cc}$, the switch from convection to overshooting is set to occur at a distance $r_0 = r_{cc} - f_0 H_{p,cc}$. $D_0$ is the value of the diffusion coefficient in the convective core at $r_0$. Increasing $f_0$ thereby increases $D_0$. The overall shape of the step overshooting is depicted in Fig. \ref{fig:DmixDiagram} (a), and is set in MESA by the two parameters: $f_0$ and $\alpha_{ov}$. To account for the step $f_0 H_{p,cc}$ taken inside the convective core, we effectively set the extent of the overshooting region as $(f_0 + \alpha_{ov})H_{p,cc}$.

\subsection{Exponential diffusive overshooting}\label{Sec:ExpOv}

Instead of just enlarging the size of the core, the exponential overshooting description assumes that the efficiency of the mixing decreases for particles further away from the convective core.  Such a decrease in mixing efficiency was motivated by \citet{Freytag1996}, whose 2D hydrodynamical simulations of surface convection in A-type stars and white dwarfs showed an exponential decay with distance from the convective boundary in the vertical velocities of the convective cells. The parameters of exponential diffusive mixing used in MESA are described by \citet{Herwig2000}, who follow the prescription of the time-dependent overshoot mixing given by \citet{Freytag1996}. \citet{Herwig2000} used this description of the overshooting to study its effect on the evolution of Asymptotic giant branch (AGB) stars, showing a clear effect on, e.g., the third dredge-up. In this sense, it concerns convective undershooting towards the interior of the star.

For an exponential overshoot, the diffusion coefficient in the overshoot region is given as 

\begin{equation}
D_{OV} = D_0 \exp \left( \frac{-2 \ \left( r - r_0\right)}{f_\text{ov} H_{p,cc} }\right).
\label{eq:Dexp}
\end{equation}

\noindent The shape of the exponential overshooting is illustrated in Fig. \ref{fig:DmixDiagram} (b). As in the case  of step overshooting, the switch from convection to overshooting is set to occur at $r_0$. To take into account the step taken inside the convective region, we effectively use $(f_0 + f_\text{ov})H_{p,cc}$ in Eq. (\ref{eq:Dexp}) instead of just $f_\text{ov} H_{p,cc}$. In MESA, the parameters $f_0$ and $f_\text{ov}$ can be varied.

\subsection{Extended exponential overshooting}\label{Sec:ExtExpMESA}

Through 2D and 3D hydrodynamical simulations of He-shell flash convection in AGB stars, \citet{Herwig2007} found that the convective boundary mixing at the bottom of the convective envelope is best described by two exponential terms. This double exponential overshooting is thereby an extension of the exponential diffusive overshooting described above and is illustrated in Fig. \ref{fig:DmixDiagram} (c). \citet{Battino2016} interpreted the mixing from the first exponential term to arise from Kelvin-Helmhotz instabilities, and the second term as being due to internal gravity waves generated at the convective boundary. The parameterised version of this extended exponential overshooting was described and applied by \citet{Battino2016} to study s-process nucleonsynthesis in AGB stars. Here, we test this description for overshooting at the core, rather than undershooting at the envelope.

As in the case of the standard exponential overshoot, $r_0$ gives the position at which the switch from convection to overshooting defined by $f_0 \cdot H_{p,cc}$ occurs, and $D_0$ is the diffusion coefficient at $r_0$. Two length scales occur: \textbf{1)} $f_1 \cdot H_{p,cc}$, which corresponds to the description in Eq. (\ref{eq:Dexp}), and \textbf{2)} $f_2 \cdot H_{p,cc}$, which takes effect for $r > r_2$. The location of $r_2$ is determined by the choice of $D_2$. In other words, when the diffusive mixing coefficient in the overshooting region decreases below $D_2$, the overshooting region is extended by a second exponential term. 

The mathematical description of the extended exponential overshooting is:\\

\noindent \underline{For $r \leq r_2$:}

\begin{align}
D_{OV} = D_0 \exp \left( \frac{-2 \ \left(r - r_0 \right)}{f_1 \cdot H_{p,cc}}\right)
\label{Eq:Dextexp1}
\end{align}

\noindent \underline{For $r \geq r_2$:}

\begin{align}
D_{OV} = D_2 \exp \left( \frac{-2 \ \left(r - r_2 \right)}{f_2 \cdot H_{p,cc}}\right).
\label{Eq:Dextexp2}
\end{align}

\noindent When using the extended exponential overshoot, the parameters to be varied in MESA are: $f_0$, $f_1$, $f_2$ and $D_2$. In all cases, it is required that $f_2 > f_1 > f_0$. If $f_2 = f_1$, one simply reproduces the single exponential overshooting. If $f_2 < f_1$, the 'extension' cuts off the single exponential overshooting and causes it to go to zero faster. To take into account the step $f_0 H_\text{p,cc}$ taken inside the convective core, we effectively use $(f_0 + f_1)H_{p,cc}$ in Eq. (\ref{Eq:Dextexp1}) and $(f_0 + f_2)H_{p,cc}$ in Eq. (\ref{Eq:Dextexp2}).

\subsection{Extra diffusive mixing in the radiative envelope}\label{Sec:EnvelopeMixing}

Aside from mixing caused by convection and convective core overshooting, additional mixing may occur in the radiative envelope. The detection of enhanced N abundance at the surface of massive stars \citep[e.g.][]{Hunter2008} is a sign of efficient chemical mixing throughout the radiative envelope. Since N is the most important byproduct of the CNO cycle during the MS and reveals itself in clear spectral lines in optical spectroscopy, it is the best element to trace envelope mixing, although O and C are also suitable diagnostics \citep{Martins2015}. The N excess of OB-type stars is typically in the range up to some 0.7 dex and has been interpreted in terms of rotational mixing \citep[e.g.][]{Brott2011}. Nevertheless, alternative explanations such as pulsational mixing have been considered as well because of a lack of correlation between rotational frequency at the stellar surface and N abundance \citep{Aerts2014}. Given that g-modes have recently shown the need of envelope mixing in addition to core overshooting \citep{Moravveji2015, Moravveji2016ApJ...823..130M}, we investigate the physical cause from a combined approach of asteroseismology and surface abundances.

The standard procedure in MESA is to include an extra constant diffusive mixing $D_\text{ext}$ in the radiative envelope as illustrated by the green shaded regions in Fig. \ref{fig:DmixDiagram} (a) and (b). The extra diffusive mixing effectively changes the shape of the chemical gradient left behind as the core contracts during the MS of stars with masses above $\sim 1.5 \ \text{M}_\odot$. Therefore, g-modes are able to trace the efficiency of such extra mixing in the near core region of the star. For the extended exponential overshooting description in Sect. \ref{Sec:ExtExpMESA}, we replace the extra constant envelope mixing by the extended exponential term (green shaded region in Fig. \ref{fig:DmixDiagram} (c)), in order to test whether or not similar g-mode pulsation patterns can be obtained when compared to the constant envelope mixing in Fig. \ref{fig:DmixDiagram} (b).

It has recently been shown by \citet{Rogers2017} that internal gravity waves (IGWs) are able to induce chemical mixing in the radiative envelope of stars with convective cores. The diffusive mixing profile resulting from IGWs increases towards the surface of the stars and is different from the constant diffusive mixing included in MESA. To investigate the effect on the g-modes and surface abundances from using such a description, we implement the diffusive mixing profile illustrated in Fig. 4 of \citet{Rogers2017} for a frequency spectrum at generated velocities $\propto \omega^{-1}$. Here $\omega$ is the angular frequency of the waves. This profile is directly loaded into MESA and rescaled to a minimum $D_\text{ext}$ at the switch from exponential overshooting to radiative envelope mixing, corresponding to the  minimum diffusive mixing set in MESA. The difference between constant mixing and the diffusive mixing profile from IGWs is illustrated in Fig. \ref{fig:DmixDiagram} (d).

\subsection{General context of 3D simulations}
The 3D hydrodynamic simuluations of core convection in a $2 \ \text{M}_\odot$ A-type star carried out by \citet{Browning2004} show that the convective boundary mixing consists of two regions. In the inner region $\nabla = \nabla_\text{ad}$ thereby affecting both chemical mixing and entropy, whereas in the outer region $\nabla = \nabla_\text{rad}$ \citep[see also][]{Viallet2015}. While the extent of the individual regions depend on the latitude, they combine to an overall spherical shape of the core. The numerical simulations of IGWs in a $3 \ \text{M}_\odot$ star also show a combination of convective penetration and overshooting at the convective core boundary, both of which decrease in depth for increasing rotation and evolve in time \citep{Rogers2013}. 3D simulations of core dynamos in B-type stars show that magnetic fields generate in the convective core also extend into the overshooting region  \citep[e.g.,][]{Augustson2016}, possibly impacting the mixing and gravity wave excitation in this region.

On the other hand, numerical simulations carried out by, e.g., \citet[][oxygen-burning shell]{Meakin2007} and \citet[][core convection simulations in a mid MS $15 \ \text{M}_\odot$ star]{Gilet2013} suggest that turbulent entrainment provides a better description for convective boundary mixing than the classical picture of convective core overshooting. This form of mixing delivers a particular parameterarisation of overshooting \citep{Viallet2015}, but does not cover the circumstances in intermediate-mass stars pulsating in high-order g-modes.

The aim of this work is not to argue which description is the best, but rather investigate whether or not period spacing series of g-modes are able to provide observational constraints on the shape of the mixing in the stellar interior. We test this for the simple prescriptions in Sect. \ref{Sec:StepOv} to \ref{Sec:EnvelopeMixing}

\section{General setup}\label{sec:GeneralSetup}

\subsection{MESA setup}

In order to test to what extent we can use g-modes to distinguish between the different shapes of convective core overshoot and radiative envelope mixing, we compute a grid of non-rotating MS models around a set of benchmark models, which are listed in Table \ref{Tab:Benchmark}. Aside from the few input parameters which are varied in the different grids, the general MESA and GYRE setup are the same (given in Appendix \ref{App:MESAinlists} and \ref{App:GYREinlist}). To carry out the model computations we use the Ledoux criterion for convection and fix the semi-convection parameter to $\alpha_\text{sc} = 0.01$. The adopted mixing length theory is the one developed by \citet{Cox1968}, and the mixing length parameter is set to $\alpha_\text{mlt} = 2.0$. We use the opacity tables from \citet{Moravveji2016MNRAS.455L..67M}, based on the \citet{Asplund2009} metal mixture and including a $75\%$ increase in the monochromatic opacities of iron and nickel from the default MESA opacity tables. This increase is motivated by the direct measurements of the iron opacities performed by \citet{Bailey2015}, and has been found to successfully explain the majority of excited modes in $\beta$ Cep and SPB stars \citep{Moravveji2016MNRAS.455L..67M}. However, none of our conclusions on the shape of the overshooting depend on the choice of opacities. The model's atmosphere is obtained from the MESA photospheric tables \citep{Paxton2011}, which are constructed from the PHOENIX \citep{Hauschildt1999a,Hauschildt1999b} and \citet{Castelli2003} model atmospheres.

In order to assure that the difference in central hydrogen content $X_c$ between two steps on the evolutionary track of the models is $\lesssim 0.001$, we set the maximum time step allowed  to be 100000 yrs. The final two parameters which we explicitly fix are $f_\text{0}$ and the initial metallicity $Z_\text{ini}$. As explained in Sect. \ref{sec:MESA3OV}, $f_\text{0}$ specifies the value of the diffusive mixing coefficient $D_\text{0}$ at the switch from convection to overshooting. To ensure $D_\text{0} > 0 \ \text{cm}^2 \ \text{s}^{-1}$, $f_\text{0} > 0$ is required. We choose $f_\text{0} = 0.001$ which corresponds to $\approx 2$ steps in resolution into the convective core from the convective boundary. 

$Z_\text{ini}$ is set to 0.014, the Galactic standard for B--type stars in the solar neighbourhood from \citet{Nieva2012} and \citet{Przybilla2013}. Through the modeling of the period spacing series of the SPB star KIC 10526294, \citet{Moravveji2015} found that an equally good model fit for a higher $M_\text{ini}$ can be obtained by reducing $Z_\text{ini}$. Therefore, we fix $Z_\text{ini}$ and vary the mass. The final MESA setup is given in Appendix \ref{App:MESAinlists}. 

\begin{table}	
	\caption{Benckmark models and their parameters.}
	\centering
	\label{Tab:Benchmark}
	\begin{tabular}{ccccccc}
	\hline\\[-1.5ex]
	\text{Benchmark}	&	\text{$M_\text{ini}$ }		& \text{$X_\text{ini}$} 		& \text{$X_\text{c}$} & \text{$\alpha_\text{ov}$} & \text{$f_\text{ov}$} & \text{$D_\text{ext}$} \\[0.5ex]
	\text{model}&\text{[M$_\odot$]}	& &	&	&	&			\text{[cm$^2$ s$^{-1}$]}	\\[0.5ex]
	\hline\\[-1.5ex]
		 \textbf{A} &	3.25	&	0.71	  &	0.50 &	-	&	0.015	&	20\\[0.5ex]
		 \textbf{B} &	3.25	&	0.71	  &	0.10 &	-	&	0.015	&	20\\[0.5ex]
		 \textbf{C} &	3.25	&	0.71	  &	0.50 &	0.15	&	-	&	20	\\[0.5ex]
		 \textbf{D} &	3.25	&	0.71	  &	0.10 &	0.15	&	-	&	20	\\[0.5ex]		 
		\hline
	\end{tabular}
	\vspace{1ex}
	
     \raggedright\small \textbf{Notes:} Benchmark model A and B (C and D) use the exponential (step) overshoot description for two different values of the central hydrogen content, $X_c$.
\end{table}

\subsection{GYRE setup}

As previously mentioned, the pulsation mode properties for the different stellar models in our computed grid are determined using the stellar oscillation code GYRE version 4.1. In this work we consider only dipole prograde g-modes, $(l,m) = (1,+1)$,  which are the easiest to detect for \emph{Kepler} SPB stars \citep{Papics2014,Papics2015,Papics2017}. 

For the GYRE computations we use the adiabatic approximation and exclude effects from rotation ($\Omega_\text{rot} = 0 \ \text{rad s}^{-1}$) on the computed frequencies. The frequency range scanned for dipole prograde modes of different radial order $n$ is set by

\begin{align}
f_\text{min} = \left( \frac{\Pi_0}{\sqrt{l(l+1)}} n_\text{max} \right)^{-1}, \qquad f_\text{max} = \left( \frac{\Pi_0}{\sqrt{l(l+1)}} n_\text{min} \right)^{-1},
\label{Eq:FreqRange}
\end{align}

\noindent where $f_\text{min}$ and $f_\text{max}$ ($n_\text{min} = 5$ and $n_\text{max} = 75$) are the minimum and maximum frequency (radial order) in the scanned range. $\Pi_0$ is the asymptotic period spacing given by $2 \pi^2\left( \int \frac{N}{r} \text{d}r\right)^{-1}$, and $N$ is the Brunt-V\"{a}is\"{a}l\"{a} frequency. By convention, GYRE uses negative values of the radial order $n$ to differentiate g-mode frequencies from p-mode frequencies. The detailed GYRE setup is given in Appendix \ref{App:GYREinlist}.

\subsection{The grids and the benchmark models}

For each of the three overshooting and extra diffusive radiative envelope mixing descriptions discussed in Sect.\ref{sec:MESA3OV}, a fine grid of stellar models is computed around two chosen central hydrogen mass fractions, $X_\text{c} = 0.5$ and $0.1$. During the early stages of main-sequence evolution (near $X_\text{c} = 0.5$) the period spacing series are very different from the ones near the terminal-age-main-sequence (TAMS, near $X_\text{c} = 0.1$). We carry out tests of the overshooting descriptions for two different values of $X_c$ to investigate if modelling a star at different ages would result in similar capacity to probe the shape of the overshooting. The benchmark models and their corresponding surrounding grids are described in detail below. 

\subsubsection{Step vs exponential diffusive overshooting grids}

\begin{figure*}
\centering
\includegraphics[width=\linewidth]{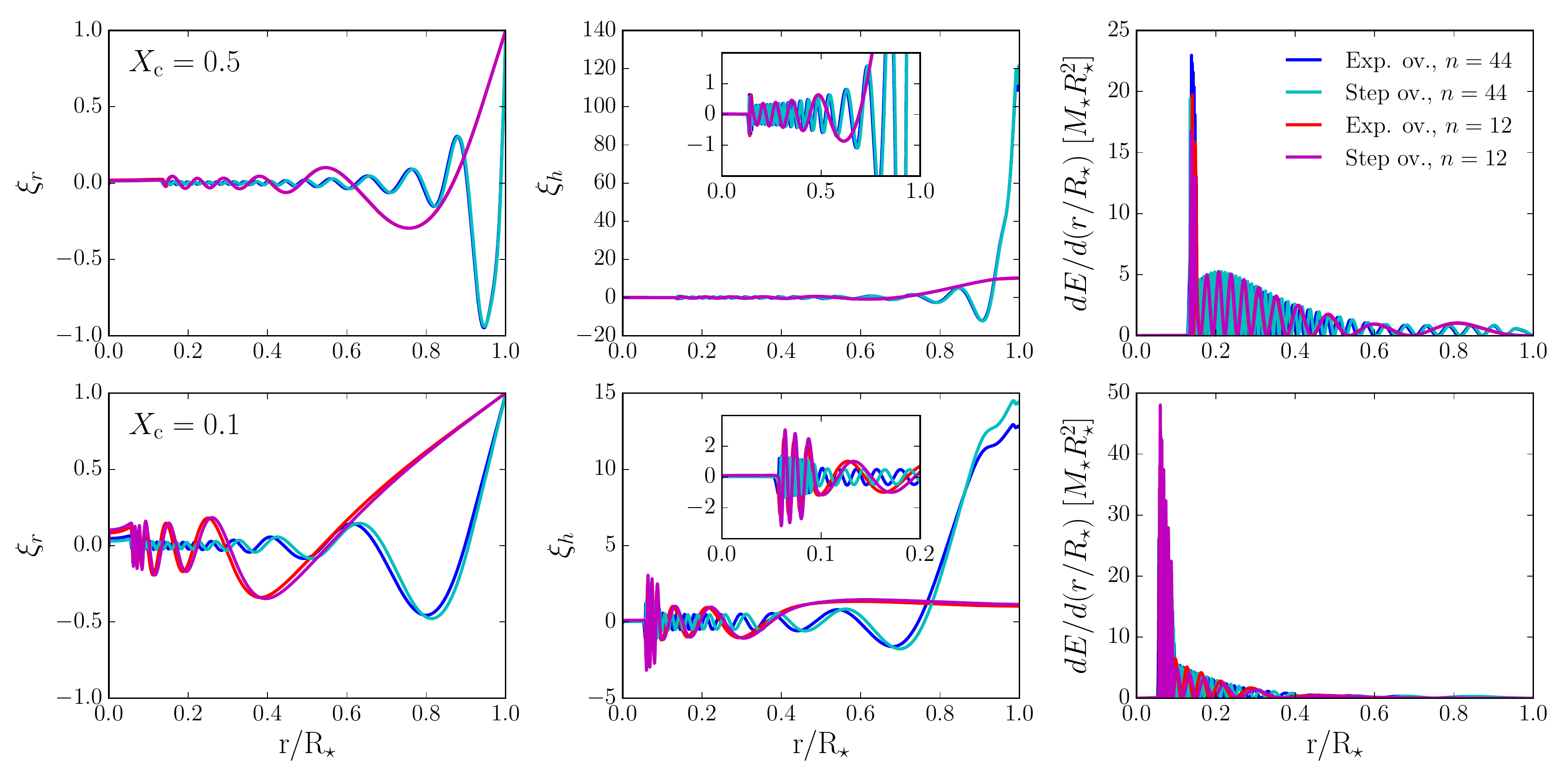}
\caption{The radial (\emph{left}) and horizontal (\emph{center}) components of the displacement vector for the $n=12$ (red colours) and $n=44$ (blue colours) dipole prograde g-modes of the benchmark models in Table \ref{Tab:Benchmark}, along with their differential mode inertia (\emph{right}). The \emph{upper} and \emph{lower} panels are for benchmark model A/C ($X_\text{c}=0.5$) and B/D ($X_\text{c}=0.1$), respectively. The dark blue and red curves are for the exponential overshoot benchmark models (A/B), whereas the cyan and magenta curves are for the step overshoot benchmark models (C/D). The components of the displacement vectors have been normalised such that the radial components are one at the surface. The inserted second plots in the center panels show a zoomed in version of the horizontal displacements.} 
\label{fig:ModeDispIntertia}
\end{figure*}

For comparing the step overshoot (Fig. \ref{fig:DmixDiagram} a) to the exponential overshoot description (Fig. \ref{fig:DmixDiagram} b), we compute two grids of stellar models and test them against the benchmark models in Table \ref{Tab:Benchmark}. For all benchmark models we set the overall abundances to the Galactic standard $(X_\text{ini}, Y_\text{ini}, Z_\text{ini}) = (0.71,0.276,0.014)$ \citep{Nieva2012,Przybilla2013}, and $M_\text{ini} = 3.25 \ \text{M}_\odot$. Benchmark model A and B (C and D) uses the exponential (step) overshoot description and only differ in age, i.e. $X_c$. As an illustration of their probing power, we show in Fig. \ref{fig:ModeDispIntertia} the prograde dipole modes of $n=12$ and $n=44$ for the four benchmark models. The period spacing series for benchmark model A and B are shown in red in Figs. \ref{fig:dPexpMFA}, \ref{fig:dPexpMFB}, \ref{fig:dPextexpMFE}, \ref{fig:dPextexpMFF}, \ref{fig:dPTamiC} and \ref{fig:dPTamiD}. 

Due to the shrinking of the convective core through the MS evolution, the mode inertia in Fig. \ref{fig:ModeDispIntertia} are shown for two different radial depths. The left and center panels demonstrates the huge difference between the radial and horizontal displacement of g-modes, with the horizontal displacement being a factor 10 to $\sim 100$ times larger than the radial displacement near the convective core for the shown radial orders. The probing power between the g-modes of different radial order is markedly different near the receding core. This property was previously exploited by \citet{Triana2015} to construct the rotation profile of the SPB star KIC 10526294.

Table \ref{Tab:GridParameters} lists the parameter setup for each of the two model grids, centered around the benchmark models A/B and C/D. For each of the grids, four parameters are varied: $M_\text{ini}$, $X_\text{ini}$, $X_c$ and the extent of the overshooting ($\alpha_\text{ov}$ for step overshoot and $f_\text{ov}$ for exponential). A given value of the $\alpha_\text{ov}$ corresponds approximately to a factor 10 lower value in $f_\text{ov}$ for B stars with $Z_\text{ini} = 0.014$ \citep{Moravveji2015}, hence the large difference in the two parameters. $D_\text{ext}$ is not varied but set to be the same as that of the benchmark models.

\begin{table}
	\centering
	\caption{Parameters for the two overshooting grids computed around benchmark models A/B and C/D.}
	\label{Tab:GridParameters}
	\begin{tabular}{ccccc}
	\hline\\[-1.5ex]
	\text{Parameter}		& \text{From} 	&	\text{To}	& \text{Step} 	&	\text{N}\\[0.5ex]
	\hline\\[-1.5ex]
	\textbf{Exponential overshoot}\\[0.5ex]
		$M_\text{ini}$ [M$_\odot$] & 3.1 & 3.4 & 0.05 &	7\\[0.5ex]
		$X_\text{ini}$ &	 0.68	& 0.73	&	0.01	 &	6\\[0.5ex]
		$X_\text{c,1}$ &		0.515	&	0.485	&	0.001	&	31\\[0.5ex]
		$X_\text{c,2}$ &		0.115	&	0.085	&	0.001	&	31\\[0.5ex]
		$f_\text{ov}$ &	0.010	&	0.020	&	0.001	&	11	\\[0.5ex]
	\hline\\[-1.5ex]
	\textbf{Step overshoot} \\[0.5ex]	
		$M_\text{ini}$ [M$_\odot$] & 3.1 & 3.4 & 0.05 &	7\\[0.5ex]
		$X_\text{ini}$ &	 0.68	& 0.74	&	0.01	 &	7\\[0.5ex]
		$X_\text{c,1}$ &		0.525	&	0.485	&	0.001	&	41\\[0.5ex]
		$X_\text{c,2}$ &		0.115	&	0.085	&	0.001	&	31\\[0.5ex]
		$\alpha_\text{ov}$ &		0.10		&	0.25		&	0.01		&	16	\\[0.5ex]	
	\hline
	\end{tabular}
	\vspace{1ex}
	
     \raggedright\small \textbf{Notes:} Start and end value, including step size and final number $N$ of the different values for a given parameter are listed. $X_\text{c,1}$ refers to the variation in central hydrogen content for benchmark models A and C, i.e. around $X_\text{c,1} = 0.5$. $X_\text{c,2}$ is the same for benchmark models B and D around $X_\text{c,2} = 0.1$.
\end{table}

\subsubsection{Grids including extra diffusive mixing in the radiative envelope}\label{Sec:DextGrid}

To test the ability of g-mode pulsations to constrain extra diffusive mixing in the radiative envelope, an additional three grids of stellar models and their pulsation properties are computed. These grids are used to carry out two comparisons. In both cases we only test the effect of varying the diffusive mixing in the radiative envelope, i.e. the effect of varying $D_\text{ext}$, $f_\text{2}$ and $D_\text{2}$, and keep all other parameters the same and constant ($M_\text{ini} = 3.25 \ \textit{M}_\odot$, $X_\text{ini} = 0.71$, $X_\text{c} = 0.5$ or $0.1$, $f_\text{ov} = 0.015$).  

\textit{The first comparison} is carried out between the exponential diffusive overshooting description including extra constant mixing in the radiative envelope (Fig. \ref{fig:DmixDiagram} (b)) and the extended exponential overshooting description (Fig. \ref{fig:DmixDiagram} (c)). For this investigation, two grids are computed and compared to benchmark model A and B in Table \ref{Tab:Benchmark}. In the first grid, the exponential overshooting description is used, and $D_\text{ext}$ is varied in the range listed in Table \ref{Tab:GridParametersDext}. When $D_\text{ext}$ is increased, the steepness of the chemical gradient resulting from the retreating convective core decreases, which in return diminishes the dips in the period spacing patterns. Keeping in mind that for the two SPB stars modeled by \citet{Moravveji2015} and \citet{Moravveji2016ApJ...823..130M}, $D_\text{ext} < 10 \ \text{cm}^{2} \text{s}^{-1}$ was needed to explain the observed period spacing series, we choose not to extend the grid to higher values.

\begin{table*}
	\centering
	\caption{Varied parameters and values for the two grids used to test the exponential overshooting with extra diffusive mixing in the radiative envelope against the extended exponential overshooting description.}
	\label{Tab:GridParametersDext}
	\begin{tabular}{ccccc}
	\hline\\[-1.5ex]
	\text{Parameter}		& \text{From} 	&	\text{To}	& \text{Step} 	&	\text{N}\\[0.5ex]
	\hline\\[-1.5ex]
	\multicolumn{3}{l}{\textbf{Exponential overshooting}}\\[0.5ex]
		$D_\text{ext}$ [cm$^2$ s$^{-1}$] &	0	&	100	&	5	&	21\\[0.5ex]
	\hline\\[-1.5ex]
	\text{Parameter}		& \multicolumn{3}{l}{\text{Full parameter range}} 	&	\text{N}\\[0.5ex]
	\hline\\[-1.5ex]
	\multicolumn{3}{l}{\textbf{Extended exponential overshooting}}\\[0.5ex]
		$f_\text{2}$ &	\multicolumn{3}{l}{[0.02, 0.05, 0.1, 0.2, 0.5, 1, 2, 5]} & 8\\[0.5ex]
		$D_\text{2}$ [cm$^2$ s$^{-1}$] &	\multicolumn{3}{l}{[1, 2, 5, 10, 20, 50, 100, 200, 500, 1000]}	&	10\\[0.5ex]
		\hline
	\end{tabular}
	\vspace{1ex}
	
     \raggedright\small \textbf{Notes:} For both grids, the parameters $M_\text{ini} = 3.25 \ \textit{M}_\odot$, $X_\text{ini} = 0.71$, $X_\text{c} = 0.5$ and $X_\text{c} =0.1$, $f = f_\text{1} = 0.015$ are held constant. In the first grid, the exponential overshooting description is used and $D_\text{ext}$ is varied. For the extended exponential overshooting, we list the values of $f_\text{2}$ and $D_\text{2}$ used for computing the grid.
\end{table*}

For the second grid, the extended exponential overshooting is used, setting $f_\text{1} = f_\text{ov}$ and $D_\text{ext} = 0 \ \text{cm}^2 \ \text{s}^{-1}$, but varying $D_\text{2}$ and $f_\text{2}$. We choose to extend $D_\text{2}$ up to $1000 \ \text{cm}^{2} \text{s}^{-1}$ and change the step size on a semi-logarithmic scale. The same is done for $f_\text{2}$. We provide the full parameter list for $f_\text{2}$ and $D_\text{2}$ in Table \ref{Tab:GridParametersDext}, rather than the minimum and maximum values. The parameter range is set in this way in order to test, if a large value of $D_\text{2}$ can mimic the period spacing series of benchmark models A and B if a similarly small value in $f_\text{2}$ is chosen. In both cases, the depth of the dips in the period spacing series become smaller and shifts towards lower periods for increasing values of $f_\text{2}$ and $D_\text{2}$, mimicking the behaviour for increasing $D_\text{ext}$. 

\textit{The second comparison} is between the exponential diffusive overshooting with constant radiative envelope mixing (Fig. \ref{fig:DmixDiagram} (b)) and with the diffusive mixing profile predicted by IGWs (Fig. \ref{fig:DmixDiagram} (d)). For both of the descriptions, the only varied parameter is $D_\text{ext}$, and the same parameter range given in Table \ref{Tab:GridParametersDext} under exponential diffusive overshooting is used. 

\subsection{The Merit Function}\label{Sec:MF}

We want to know whether g-mode pulsations can distinguish between different overshooting and envelope mixing descriptions. To do this, we define a \textit{Merit Function (MF)} which we use to rank the grid models according to how well they can mimic the benchmark

\begin{equation}
MF = \frac{1}{\left( N - k\right)\sigma_R^2} \sum_{i = 1}^N \left( f_i^{(bench)} - f_i^{(model)}\right)^2. 
	\label{Eq:MF}
\end{equation}

\noindent Here $N$ is the number of dipole prograde modes for the benchmark model in the period range $0.8-3 \ \text{d}$ for which g-modes are detected in SPB stars \citep{Papics2017}, $k$ is the number of varied parameters in the grid ($k = 4$ for both grids in Table \ref{Tab:GridParameters}, $k=1$ and 2 for the exponential and extended exponential overshooting grid in Table \ref{Tab:GridParametersDext}, respectively). $\sigma_R$ is the Rayleigh limit for the nominal \textit{Kepler} mission, $\sigma_R = 1/T = 0.00068 \ \text{d}^{-1}$. The Rayleigh limit gives the frequency resolution, and is an upper limit for the frequency error of observed pulsations. $f_i^{(bench)}$ is the \textit{i}'th benchmark model frequency within the range in period of $0.8-3 \ \text{d}$, and $f_i^{(model)}$ is the nearest frequency to the $f_i^{(bench)}$ for a given model in the grid. From this definition, a lower $MF$ corresponds to a better match, and a perfect match will return $MF = 0$. 

While the merit function ranks the grid models according to which matches the benchmark models best, it provides no information on how different the MF values of two models have to be in order to be distinguishable. Having this information is needed in order to \textbf{a)} tell which model parameter range essentially returns the same best matching model result, and \textbf{b)} if the different shapes of overshooting and radiative envelope mixing are distinguishable according to the resulting g-mode properties. Therefore, for the merit function of the best matching model MF$_\text{Best}$, we determine a 'cut-off' in MF below which the models are considered indistinguishable. To define this cut-off $f_i^{(model)}$ is replaced by $f_i^{(model)} + \delta_i$ in Eq. (\ref{Eq:MF}), where $\delta_i$ is an expected observational error estimate on the frequencies drawn randomly from a normal distribution centered around zero and with a standard deviation equal to $\sigma_R$. This step is repeated 10000 times, resulting in a normal distribution of MF values (MF$_\text{Best,10000}$), centered around MF$_\text{Best} + N/(N-k)$. This distribution represents how much MF$_\text{Best}$ would change if we shift the frequencies of the best matching model within the expected observational errors on the frequencies from the \emph{Kepler} mission. Based on this distribution, the cut-off is set to be mean(MF$_\text{Best,10000}$) + $\sigma_\text{std}$(MF$_\text{Best,10000}$) $ = \text{MF}_\text{cut}$. All models with MF below this cut-off essentially match the benchmark model just as well as the model with the lowest MF within the same grid, considering the frequency resolution of the nominal \emph{Kepler} mission.

\section{Probing power of g-modes to unravel mixing shapes}\label{Sec:ShapeComparisons}

\subsection{Step- vs. exponential overshoot}\label{Sec:StepvsExp}

We turn towards answering the question of whether or not it is possible to distinguish between step and exponential overshooting using g-modes. For this to be the case, the merit functions of the best matching models resulting from comparing benchmark model A and B to the grid of models with step overshooting have to be higher than the merit function cut-offs MF$_\text{cut,A} = 1.41$ and MF$_\text{cut,B} = 1.36$. An internal comparison between the grid models is discussed in Appendix \ref{App:Grid}.

\begin{figure*}
\centering
\includegraphics[width=0.9\linewidth]{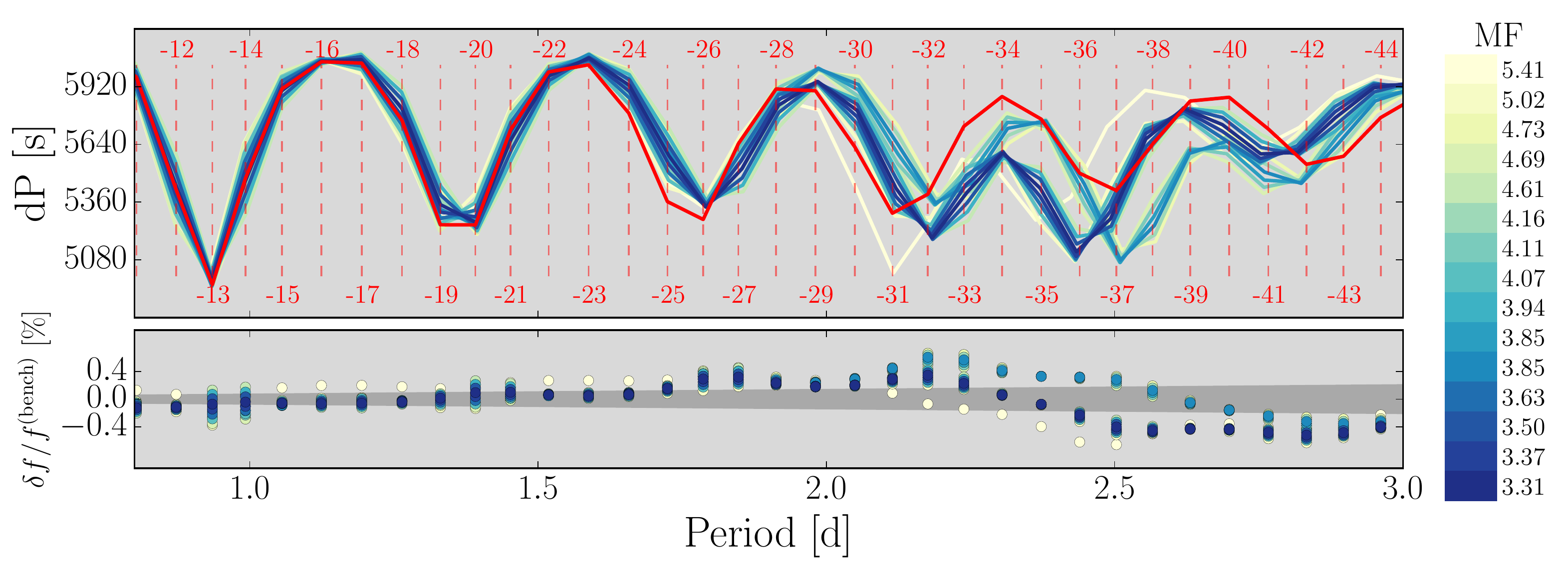}
\caption{Period spacing series (upper panel) of the 15 best matching models resulting from comparing the exponential diffusive overshoot benchmark model A (red curve) to the step overshooting grid. The colours indicate the different values of the merit function MF, and the vertical dashed lines give the positions and values of the radial orders $n$ of the benchmark model. The lower panel shows the frequency deviations in percentage between the benchmark model and the nearest frequencies of the exponential overshooting models. The position of the deviations in period space has been fixed to those of the benchmark model. The dark grey shaded region shows the Rayleigh limit for the \emph{Kepler} mission.}
\label{fig:dPexpMFA}
\end{figure*}

\begin{figure*}
\centering
\includegraphics[width=0.9\linewidth]{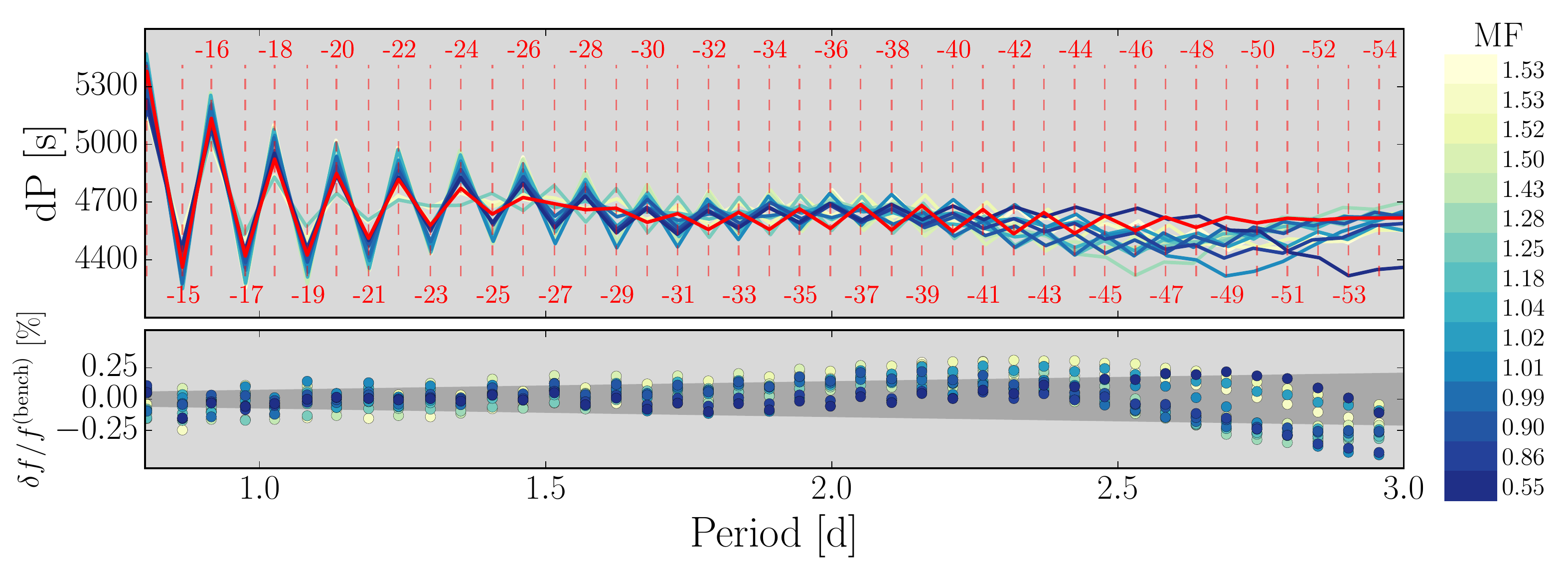}
\caption{Same as Fig. \ref{fig:dPexpMFA} but for benchmark model B compared to the step overshooting grid.}
\label{fig:dPexpMFB}
\end{figure*}

Figures \ref{fig:dPexpMFA} and \ref{fig:dPexpMFB} shows the period spacing series of the $(l,m) = (1,1)$ g-modes for the 15 best matching models obtained when comparing benchmark model A and B (red curves) to the step overshooting grid, respectively. The colours correspond to different values of the merit function, which become darker for decreasing MF values. The red dashed lines and numbers denote the positions and values of the different radial orders $n$. The lower panels illustrate the frequency deviations $\delta f / f^{(bench)}$ in percentage between the $(l,m) = (1,1)$ g-modes of the best matching models and the benchmark model. The colors link the differences to the period spacing series in the upper panels and the periods have been fixed to the ones of the benchmark models. The dark grey shaded region shows the Rayleigh limit for the nominal \textit{Kepler} mission. 

As seen in both Figs. \ref{fig:dPexpMFA} and \ref{fig:dPexpMFB}, the step overshooting models manage to reproduce well the period spacing series of the exponential overshoot benchmark model at low periods, but are unable to do so at longer periods. The best step overshooting model mimicking the g-mode frequencies of benchmark model A returns $\text{MF} = 3.31$, well above the $\text{MF}_\text{cut,A} = 1.41$. However, we find for the more evolved model comparison between benchmark model B and the step overshooting grid that the best matching model has MF = 0.55, below the cutoff MF$_\text{cut,B} = 1.36$. Therefore, we conclude that \textit{the ability of dipole prograde g-modes covering the range in period of 0.8 to 3 d to distinguish between step and exponential overshooting depends on the MS evolutionary stage. At $X_\text{c} = 0.5$ we can distinguish between the two overshooting descriptions at the level of $6\sigma_{std}$, but at $X_\text{c} = 0.1$ the step and exponential diffusive overshooting are indistinguishable within $3\sigma_{std}$.}

\subsubsection{Correlations}

\begin{figure}
\centering
\includegraphics[width=\linewidth]{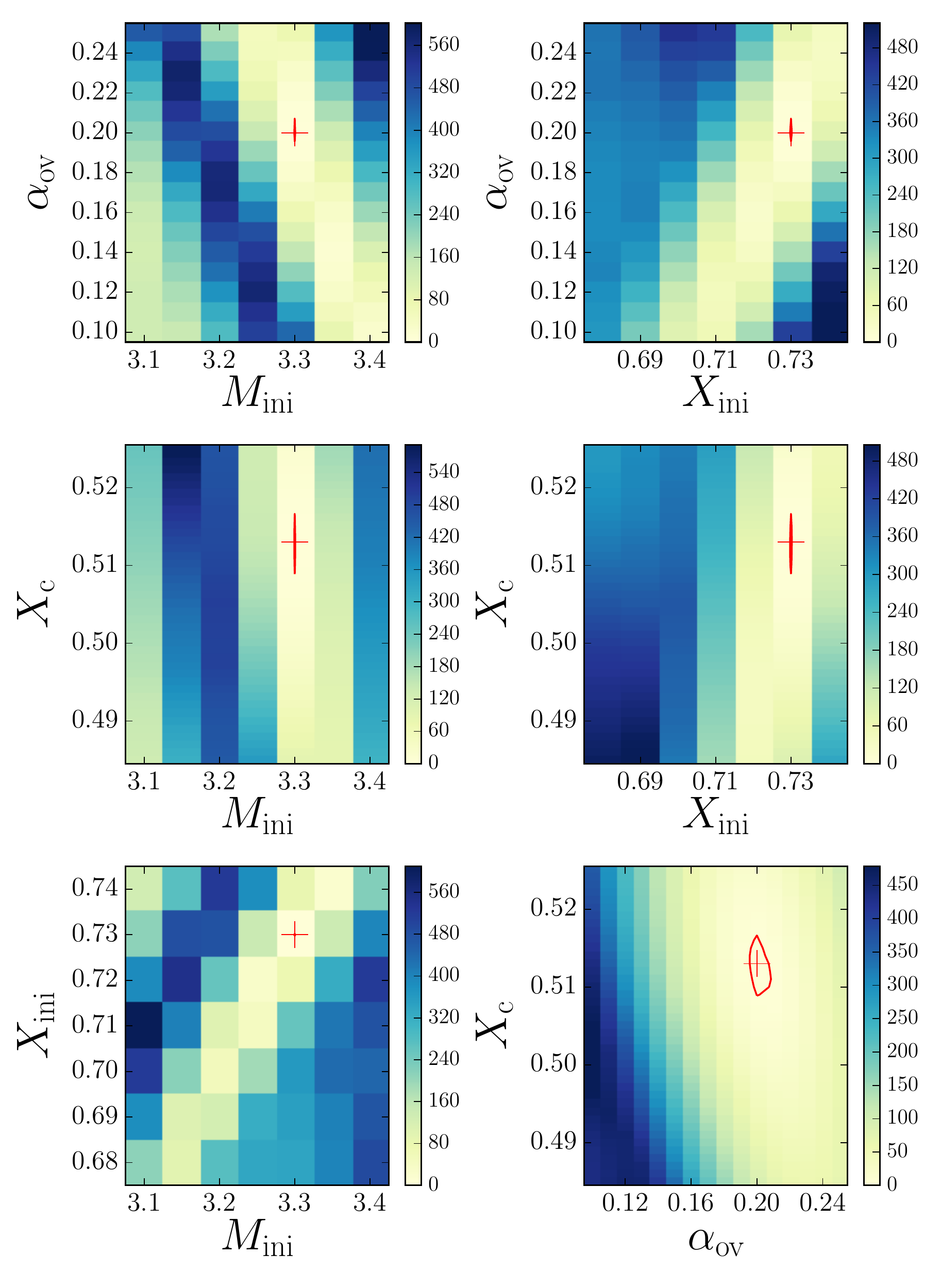}
\caption{2D plots for benchmark model A compared to the step overshooting grid, showing the change in merit function as two parameters are varied and all others are kept fixed to those of the best matching step overshooting model at $X_\text{c} \sim 0.5$. The red cross shows the position of the best fitting model in each of the plots and red curves the 5.15 merit function cut-off contour lines for the best matching step overshooting model.}
\label{fig:CorrelationsAexp}
\end{figure}

\begin{figure}
\centering
\includegraphics[width=\linewidth]{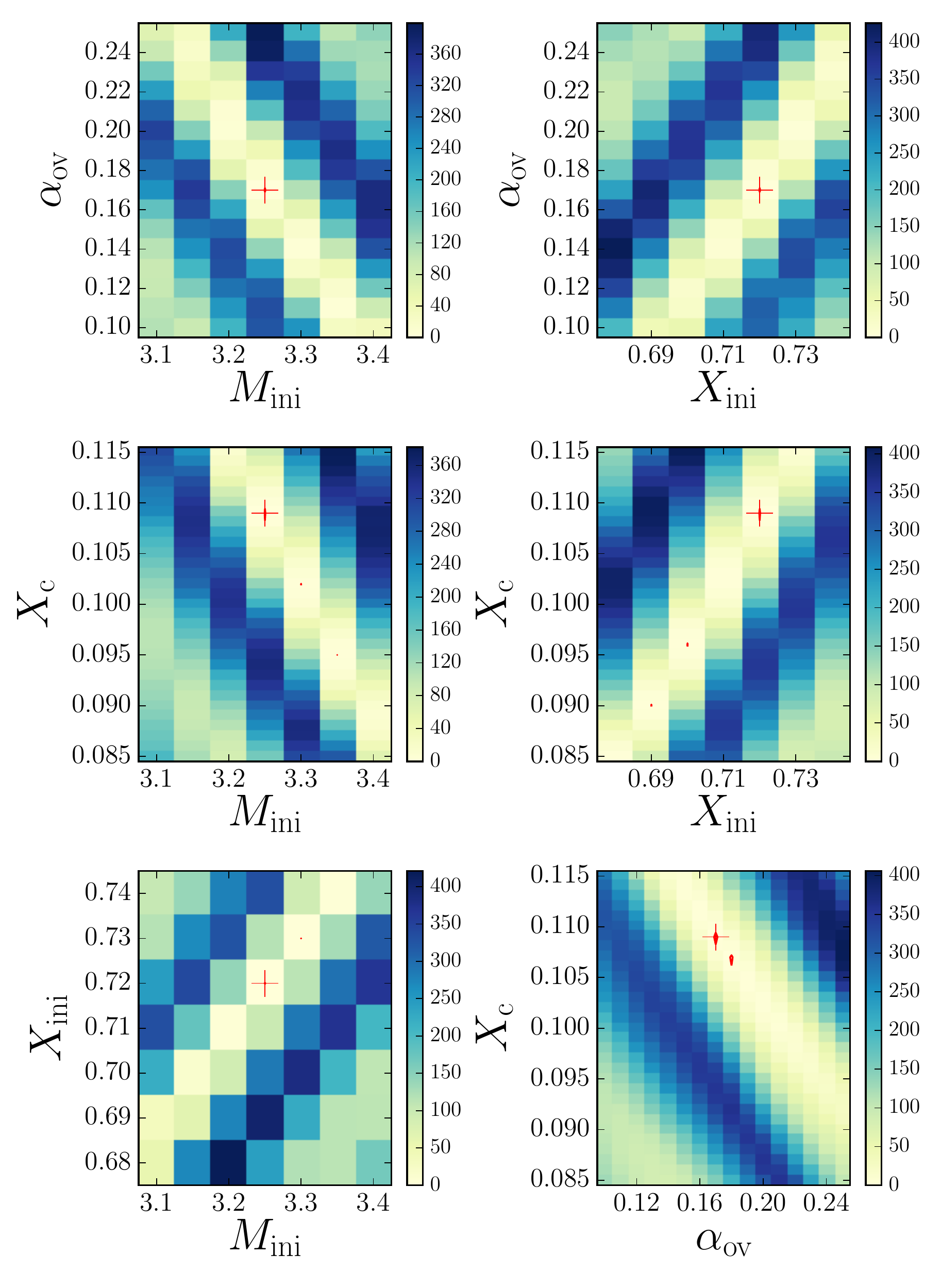}
\caption{Same as Fig. \ref{fig:CorrelationsAexp} but for the comparison between benchmark model B and the step overshooting grid at $X_\text{c} \sim 0.1$ and merit function cut-off at 2.00.}
\label{fig:CorrelationsBexp}
\end{figure}

Forward modelling of SPBs revealed correlations among the parameters, but these have not yet been analysed in detail. To search for correlations between the model parameters listed in Table \ref{Tab:GridParameters}, we first create 2D surface plots showing how the merit function varies as a function of varying two parameters at a time, while keeping all others fixed to those of the best matching step overshooting models. If any clear correlations exists between the parameters, those should show up in these 2D surface plots. The step overshooting model which best mimics the $(l,m) = (1,1)$ g-mode frequencies of benchmark model A has the parameters $M_\text{ini} = 3.30 \ \text{M}_\odot$, $X_\text{ini} = 0.73$, $X_\text{c} = 0.513$ and $\alpha_\text{ov} = 0.20$. The correlation plots resulting from centering on these model parameters are shown in Fig. \ref{fig:CorrelationsAexp}. Similarly for benchmark model B we get $M_\text{ini} = 3.25 \ \text{M}_\odot$, $X_\text{ini} = 0.72$, $X_\text{c} = 0.109$ and $\alpha_\text{ov} = 0.17$ as the best matching step overshooting model and Fig. \ref{fig:CorrelationsBexp}.

For the less evolved models, i.e. at $X_\text{c} \sim 0.5$ (see Fig. \ref{fig:CorrelationsAexp}), when $X_\text{ini}$  is increased the lowest merit function is obtained by increasing $M_\text{ini}$, $X_\text{c}$ and the extent of the overshooting $\alpha_\text{ov}$. The opposite is seen when comparing the initial mass to the extent of the overshooting, where a higher $M_\text{ini}$ requires a lower $\alpha_\text{ov}$ for a given benchmark model comparison. A similar but less clear correlation might be present between $M_\text{ini}$ and $X_\text{c}$, while no clear correlation is seen for the $X_\text{c}$ vs $\alpha_\text{ov}$ comparison. For the more evolved models at $X_\text{c} \sim 0.1$ in Fig. \ref{fig:CorrelationsBexp}, the correlations become sharper and more defined compared to the higher $X_\text{c}$ counterparts in Fig. \ref{fig:CorrelationsAexp}. Furthermore, a clear correlation is now seen for the $X_\text{c}$ vs $M_\text{ini}$ and $\alpha_\text{ov}$ comparisons, where lower merit function values are obtained when $X_\text{c}$ decreases and $M_\text{ini}$ or $f_\text{ov}$ are increased.

In order to come up with mathematical expressions for the correlations seen in Figs. \ref{fig:CorrelationsAexp} and \ref{fig:CorrelationsBexp} we carry out a multivariate linear regression of the form 

\begin{equation}
Y_{i} = \beta_{0} + \beta_{1} X_{i1} + ... +  \beta_{k} X_{ik} , 
\label{Eq:MLR}
\end{equation}

\noindent using $1/\text{MF}_i$ as weights. Here $i$ represents the $i$'th model in the grid, and $k$ the total number of compared parameters. As an example, to determine the dependence of $M_\text{ini}$ on $X_\text{ini}$, $X_\text{c}$ and $\alpha_\text{ov}$ we set $M_\text{ini} = Y$ and the predictors ($X_1$,$X_2$,$X_3$) = ($X_\text{ini}$, $X_\text{c}$, $\alpha_\text{ov}$) and estimate the regression coefficients $\beta_0$, $\beta_1$, $\beta_2$ and $\beta_3$. The significance of the predictors is defined at the conventional 5\% level, corresponding to a $p$-value $ < 0.05$. In the case that a predictor obtains a $p$-value $ \geq 0.05$, the predictor is excluded from Eq. \ref{Eq:MLR} and the multivariate linear regression is carried out once more until all predictors have $p < 0.05$.

\begin{table*}
	\centering
	\caption{Results from multivariate linear regression.}
	\label{Tab:lm}
	\begin{tabular}{ccccccc}
	\hline\\[-1.5ex]
	\text{} 	&	\text{$M_\text{ini}$}	& \text{$X_\text{ini}$} & \text{$X_\text{c}$} & \text{$\alpha_\text{ov}$} & \text{Intercept}	&	\text{$R^2$}\\[0.5ex]
	\hline\\[-1.5ex]
	\multicolumn{6}{l}{\textbf{Benchmark model A vs step overshooting grid}} \\[0.5ex]
		$M_\text{ini}$ & 	-	&	1.35$\pm$0.02	&	0.52$\pm$0.04	&	-0.35$\pm$0.01	&	2.10$\pm$0.02	&	0.134\\[0.5ex]
		$X_\text{ini}$ &		0.066$\pm$0.001	&	-	&	0.253$\pm$0.008	&	0.023$\pm$0.002	 &	0.367$\pm$0.005	&	0.123\\[0.5ex]
		$X_\text{c}$ &	0.010$\pm$0.0008	&	0.107$\pm$0.004	&	-	&	-	&	0.394$\pm$0.003	&	0.043	\\ [0.5ex]
		$\alpha_\text{ov}$	&	-0.104$\pm$0.003	&	0.15$\pm$0.01	&	-	&	-	&	0.41$\pm$0.01	&	0.036\\
		\hline\\[-1.5ex]
		\multicolumn{6}{l}{\textbf{Benchmark model B vs step overshooting grid}} \\[0.5ex]
		$M_\text{ini}$ & 	-	&	1.60$\pm$0.03	&	-1.91$\pm$0.06	&	-0.59$\pm$0.01	&	2.41$\pm$0.02	&	0.166\\[0.5ex]
		$X_\text{ini}$ &		0.071$\pm$0.001	&	-	&	0.68$\pm$0.01	&	0.084$\pm$0.003	&	0.397$\pm$0.005	&	0.188\\[0.5ex]
		$X_\text{c}$ &	-0.0188$\pm$0.0006	&	0.150$\pm$0.003	&	-	&	-0.024$\pm$0.001	&	0.058$\pm$0.002	&	0.111\\ [0.5ex]
		$\alpha_\text{ov}$	&	-0.130$\pm$0.003	&	0.42$\pm$0.01	&	-0.54$\pm$0.03	&	-	&	0.35$\pm$0.01	&	0.088\\
		\hline\\[-1.5ex]
	\end{tabular}
	\vspace{1ex}
	
     \raggedright\small \textbf{Notes:} Included are the \textit{parameter estimates $\pm$ standard errors} of the predictors for which $p < 0.05$. $R^2$ the fraction of the variance explained by the included predictors and their coefficients.
\end{table*}

Table \ref{Tab:lm} lists the results for the benchmark model A and B comparison to the step overshooting grids. Each row corresponds to an expression of the form in Eq. (\ref{Eq:MLR}), and the elements are the estimated $\beta_k$ and their standard errors. For all the predictors with listed coefficients, the $p$-value is $\leq 0.0001$. The results in Table \ref{Tab:lm} and Figs. \ref{fig:CorrelationsAexp}, \ref{fig:CorrelationsBexp} reveal that correlations between the basic parameters used in forward modelling may be strong but change over time along the evolutionary track. Our results offer a useful guide to refine forward modelling from g-modes once rough $X_\text{c}$-values have been found.

\subsection{Extra constant diffusive mixing vs extended exponential overshoot}\label{Sec:DextvsExtExp}

Like in Sect. \ref{Sec:StepvsExp} and Appendix \ref{App:Grid} we first determine how benchmark model A and B compare intrinsically to the grid of models with exponential overshooting and extra constant diffusive mixing in the radiative envelope (Fig. \ref{fig:DmixDiagram} b) in Table \ref{Tab:GridParametersDext}. For benchmark model A, we find that exponential diffusive overshooting models with $D_\text{ext} = 15-25 \ \text{cm}^2 \ \text{s}^{-1}$ match benchmark model A equally well. At a lower $X_\text{c}$ closer to the TAMS, this range increases to $D_\text{ext} = 15-35 \ \text{cm}^2 \ \text{s}^{-1}$ for benchmark model B.

\begin{figure*}
\centering
\includegraphics[width=0.9\linewidth]{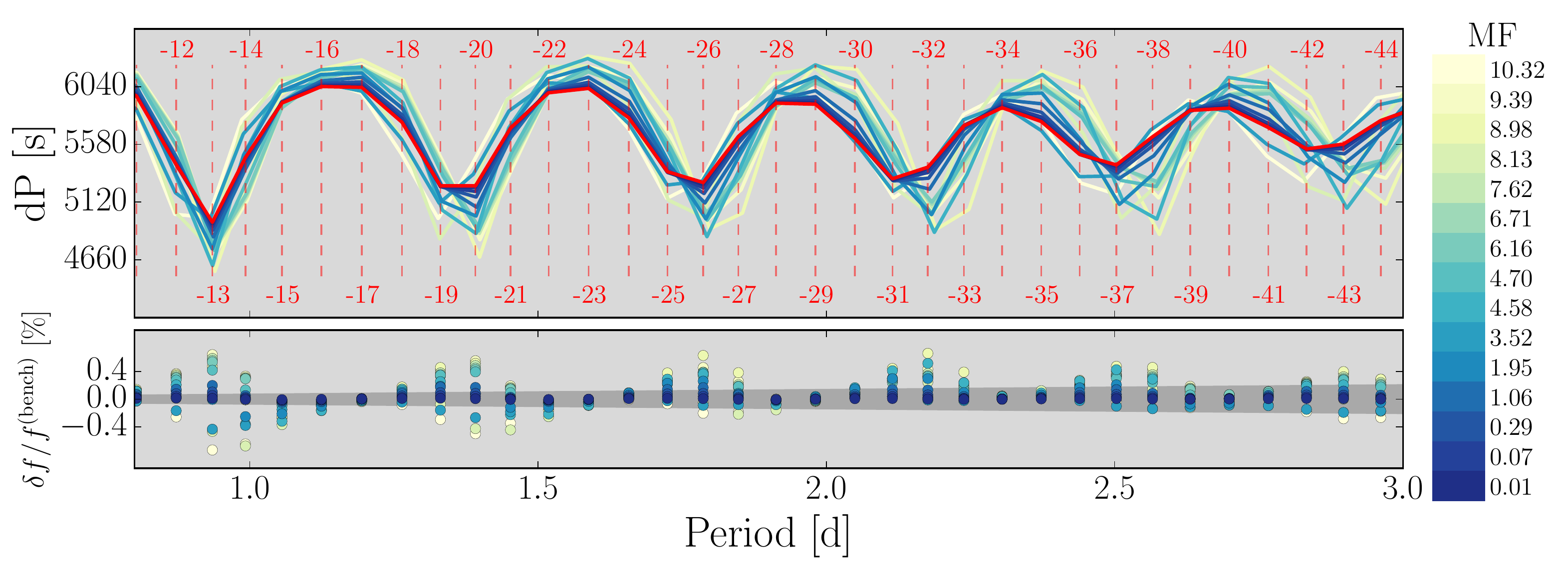}
\caption{Period spacing series and frequency deviations for the 15 extended exponential overshooting models which mimic the g-mode pulsations of benchmark model A (red curve) the best. For further details see text and Fig. \ref{fig:dPexpMFA}.}
\label{fig:dPextexpMFE}
\end{figure*}

\begin{figure*}
\centering
\includegraphics[width=0.9\linewidth]{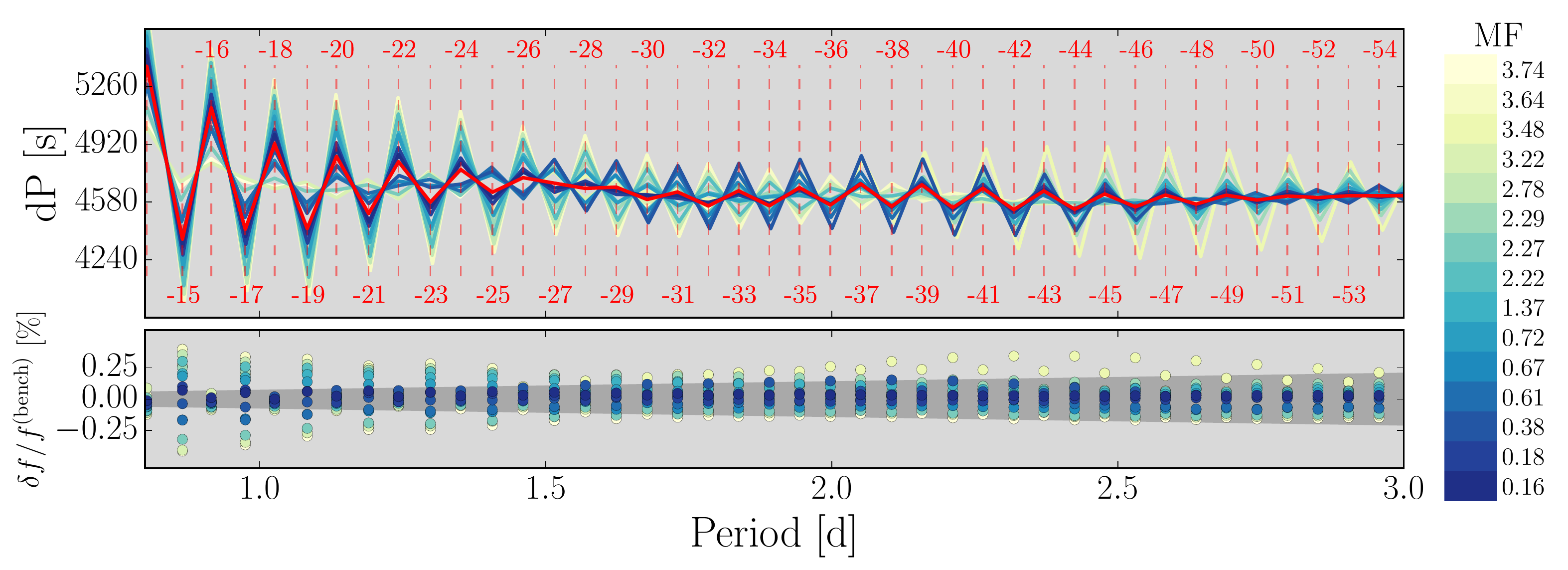}
\caption{Same as Fig. \ref{fig:dPextexpMFE} but for benchmark model B.}
\label{fig:dPextexpMFF}
\end{figure*}

\begin{figure*}
\centering
\includegraphics[width=0.9\linewidth]{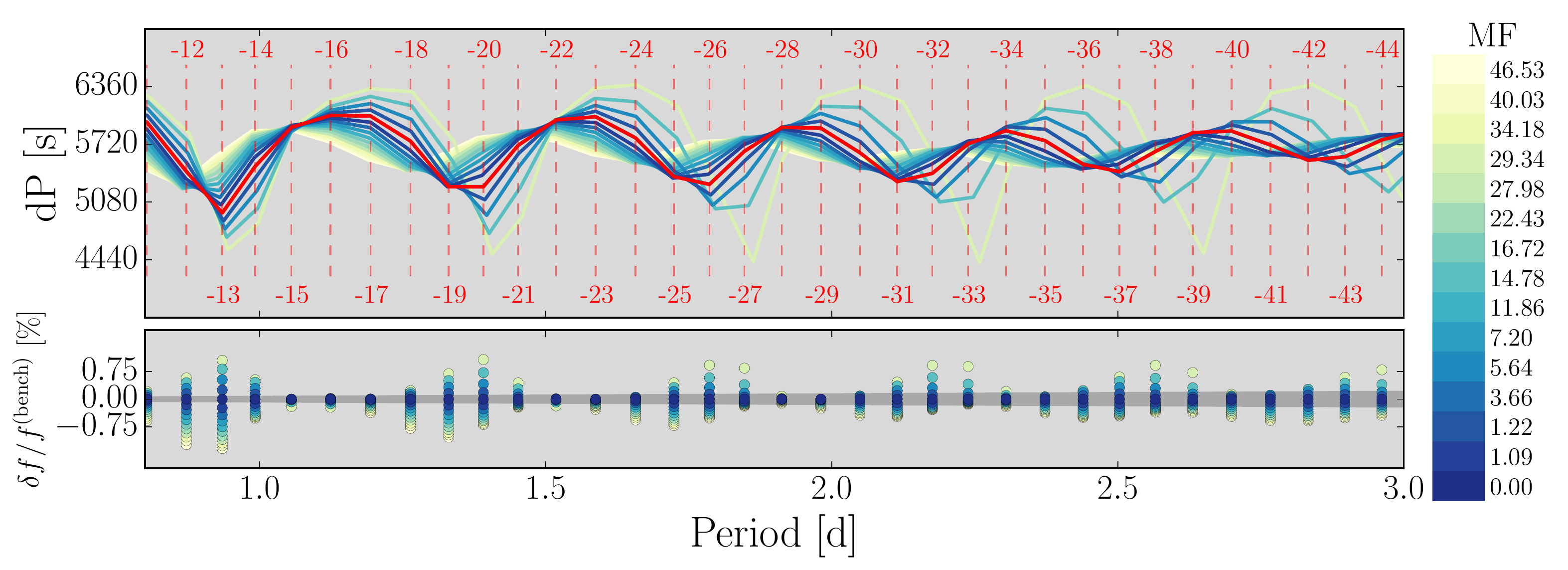}
\caption{Period spacing series and frequency deviations for the 15 exponential diffusive overshooting models with the increasing radiative envelope mixing profile of the IGWs which mimic the g-mode pulsations of benchmark model A (red curve) the best. For further details see text and Fig. \ref{fig:dPexpMFA}.}
\label{fig:dPTamiC}
\end{figure*}

\begin{figure*}
\centering
\includegraphics[width=0.9\linewidth]{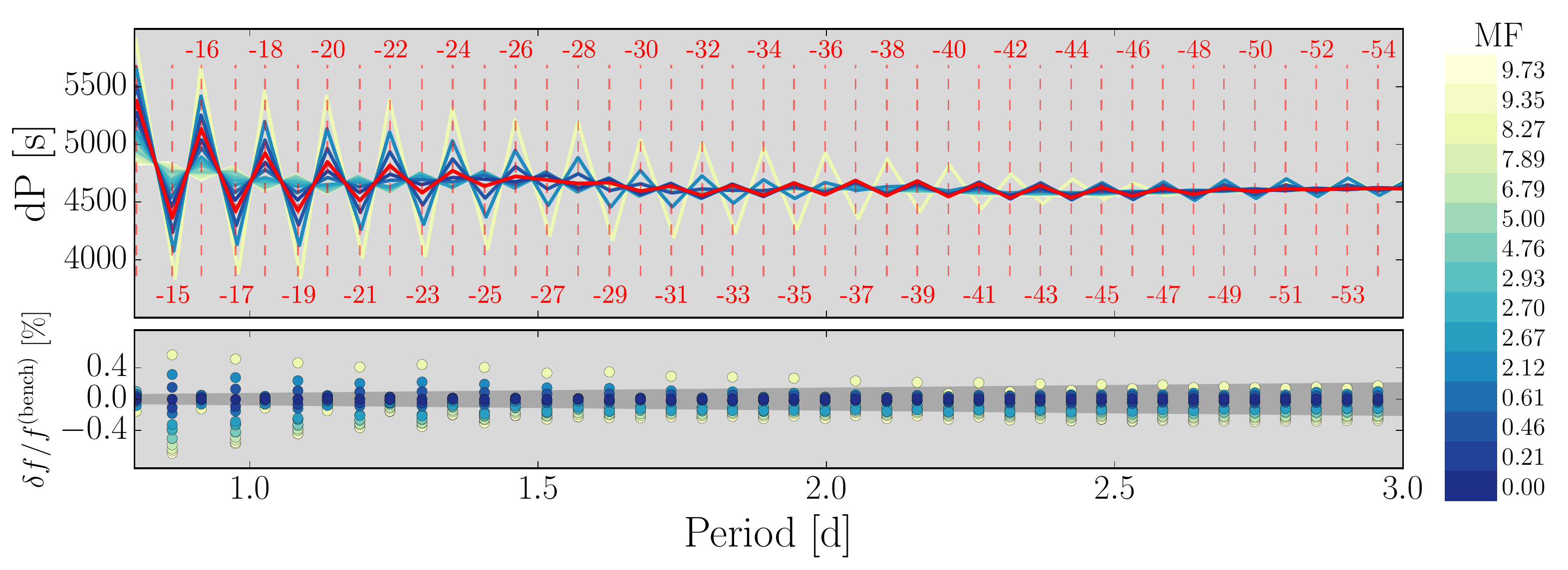}
\caption{Same as Fig. \ref{fig:dPTamiC} but for benchmark model B.}
\label{fig:dPTamiD}
\end{figure*}


The period spacing series of the 15 best matching models resulting from comparing the $(l,m) = (1,1)$ g-mode frequencies of benchmark model A and B to the grid of extended exponential overshooting models (i.e. Fig. \ref{fig:DmixDiagram}b vs Fig. \ref{fig:DmixDiagram}c) listed in Table \ref{Tab:GridParametersDext} are shown in Figs. \ref{fig:dPextexpMFE} and \ref{fig:dPextexpMFF}. For both figures the period spacing series of all 15 models mimic the one of the benchmark models much better than what is seen in Figs. \ref{fig:dPexpMFA} and \ref{fig:dPexpMFB}. This is also indicated by the increasing number of frequency deviations falling within the dark grey band of the Rayleigh limit. For Fig. \ref{fig:dPextexpMFF} in particular, the majority of the frequency deviations for periods above $\sim 2 \ \text{d}$ fall within the Rayleigh limit. Therefore, the determination of the best matching model is mostly depending on the g--mode pulsations with periods below $1.5 \ \text{d}$. This is less so the case for $X_\text{c} = 0.5$, where the frequency deviations in most cases still fall outside the Rayleigh up to periods of $\sim 2.5 \ \text{d}$.

The extended exponential overshooting model which best mimics the g-modes of benchmark model A has MF$ = 0.01$. This is $5\sigma_{std}$ below the $\text{MF}_\text{cut,A} = 1.41$ cutoff, whereas the model with the lowest merit function (MF$= 0.16$) for Xc = 0.1 (benchmark model B) is more than $4\sigma_{std}$ below $\text{MF}_\text{cut,B} = 1.36$. Based on these results, we conclude that \textit{it is not possible to distinguish between g-mode pulsations for models with exponential overshooting including extra constant diffusive mixing in the radiative envelope and the extended exponential overshooting description at either $X_\text{c} = 0.5$ or $0.1$.}

Finally, we find that at $X_\text{c} = 0.5$ setting $D_\text{2} = D_\text{ext}$ and $f_\text{2} \geq 0.5$ results in identical period spacing series within estimated observational frequency errors. At a lower $X_\text{c}$, however, the span in $D_\text{2}$ values becomes larger (20 to $100 \ \text{cm}^2 s^{-1}$), and a higher $D_2$ generally requires a lower $f_2$. At both $X_\text{c} = 0.5$ and 0.1 the best matching extended exponential overshooting model has $D_2 = 20 \ \text{cm}^2 s^{-1}$ and $f_2 = 5.0$.

\subsection{Extra constant diffusive mixing vs chemically induced mixing from IGWs}\label{Sec:DextvsTami}

As a next test, we investigate whether or not g-modes are affected at a distinguishable level if we use a mixing profile obtained from 2D hydrodynamical simulations of IGWs \citep{Rogers2017} instead of a constant diffusive mixing in the radiative envelope (mixing profiles in Figs. \ref{fig:DmixDiagram} b and d). For a discussion and review on additional radiative envelope mixing processes, such as rotation, we refer the reader to \citet{Zahn2011} and \citet{Mathis2013}. We use the exponential diffusive overshooting description for both the benchmark and grid models, fixing all other parameters except from $D_\text{ext}$. 

Figures \ref{fig:dPTamiC} and \ref{fig:dPTamiD} shows the 15 best matching models resulting from comparing benchmark model A and B with constant mixing in the radiative envelope to the grid of models for which the mixing profile of the IGWs from \citet{Rogers2017} has been implemented in MESA. The period spacing series and frequency deviations for these models are very similar to the ones seen in Figs. \ref{fig:dPextexpMFE} and \ref{fig:dPextexpMFF} for the extended exponential overshooting grid. Once again the grid models which mimic the g-modes of the benchmark model best have merit functions below the $\text{MF}_\text{cut,A} = 1.41$ and $\text{MF}_\text{cut,B} = 1.36$ cut-offs. In fact in both cases MF$_\text{Best} = 0.00$. We conclude that \textit{within more than $5\sigma_{std}$ significance, $(l,m) = (1,1)$ g-modes behave the same for a constant mixing throughout the envelope as for chemical mixing profiles from IGWs at both $X_\text{c} = 0.5$ and $0.1$.} These results give a full justification of the forward modelling strategy adopted by \citet{Moravveji2015,Moravveji2016ApJ...823..130M} to deduce the best value for $D_\text{ext}$. Very low values for $D_\text{ext}$ were found in this way, in contrast to theoretical predictions \citep{Mathis2004} or numerical simulations \citep{Prat2016} on mixing induced by vertical shear instability. This discrepancy remains to be understood after several more stars will have been modelled seismically.

\section{Combined probing power of g-modes and surface abundances}\label{Sec:gmodesAndN14}

While we have shown that g-modes alone are not sufficient to constrain the mixing throughout the entire envelope but mainly $D_\text{ext}$ in the near-core region, combining information from these pulsations with expected enhanced surface abundances of N$^{14}$ ($= 12 + \log[\text{N}^{14}/\text{H}^{1}]$) might be. The efficiency of the radiative envelope mixing increases towards the surface in Fig. \ref{fig:DmixDiagram} d for the mixing due to IGWs. This implies that a much lower mixing near the overshooting region can transport more  nitrogen produced in the CNO cycle to the surface of the star compared to the case of a constant radiative envelope mixing. Therefore, if period spacing patterns are detected in combination with enhanced N$^{14}$ abundances, this would give a way to deduce the mixing profile from the overshoot region to the surface. To test this, we expand our grid of stellar models with the radiative envelope mixing profile from \citet{Rogers2017} with starting values $D_\text{ext} = 500$ to $19000 \ \text{cm}^2 \ \text{s}^{-1}$.

\begin{figure}
\centering
\includegraphics[width=\linewidth]{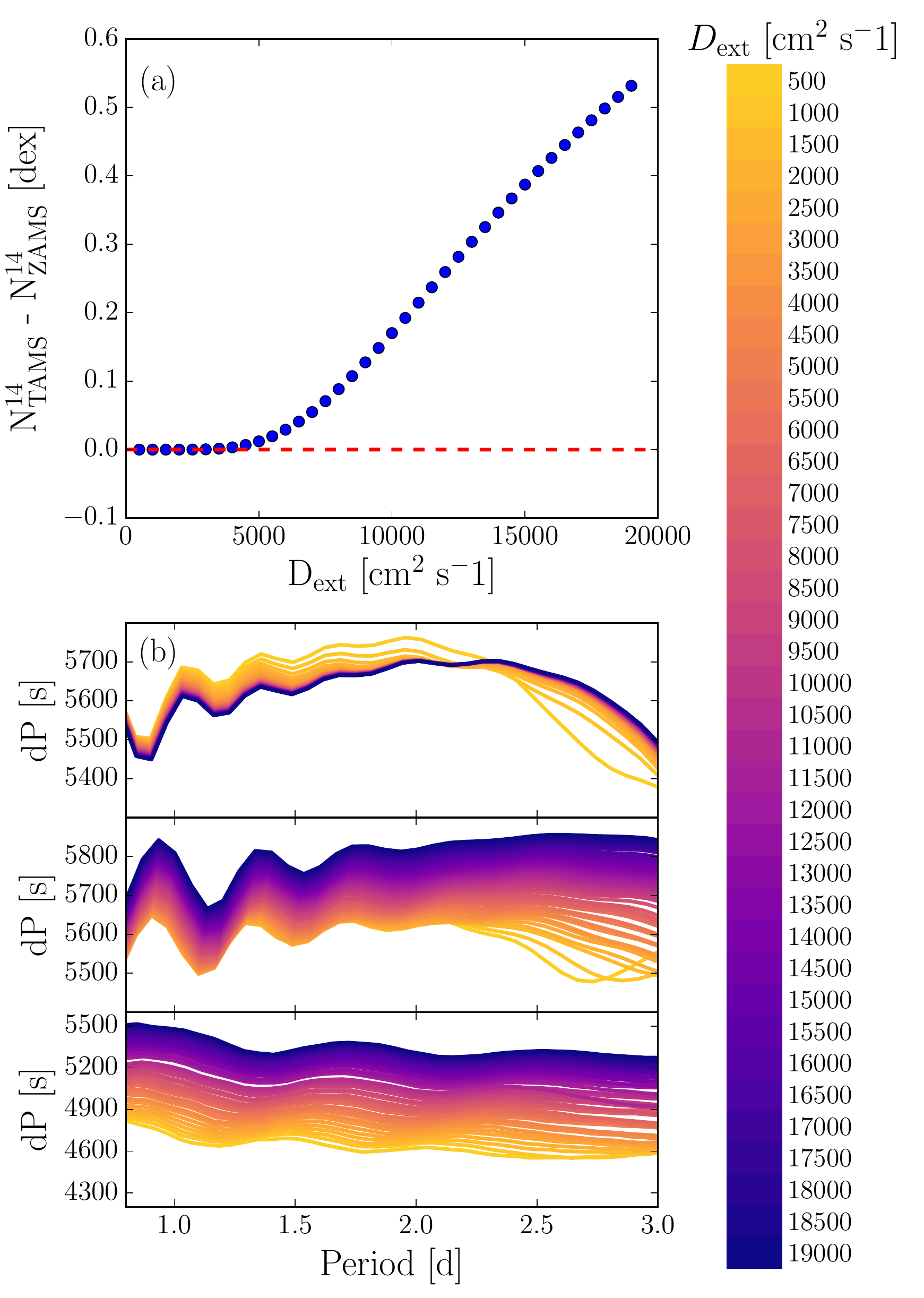}
\caption{Surface nitrogen abundance enhancement during MS evolution as a function of $D_\text{ext}$ (a) for an increasing radiative envelope mixing towards the surface of the stellar model based on the $D_\text{mix}$ profile in Fig. \ref{fig:DmixDiagram} d. The red dashed line shows the change in N$^{14}$ for a constant envelope mixing (Fig. \ref{fig:DmixDiagram} b). b) shows the corresponding effects of varying $D_\text{ext}$ on the $(l,m) = (1,1)$ period spacing series at $X_\text{c} = 0.7$, 0.5 and 0.1 (top, middle and lower panel respectively). See text for further explanation.}
\label{fig:Abundance}
\end{figure}

In Fig. \ref{fig:Abundance} a, the difference in surface abundance of nitrogen from ZAMS to TAMS (for which we use $X_\text{c} = 0.01$) is plotted as a function of $D_\text{ext}$ for the radiative envelope mixing in Fig. \ref{fig:DmixDiagram} d. For $D_\text{ext} \gtrsim 5000 \ \text{cm}^2 \ \text{s}^{-1}$ N$^{14}$ starts to be transported to the surface of the star within the MS life-time. In comparison, no difference in the surface abundance of nitrogen between the ZAMS and TAMS is seen for a constant radiative envelope mixing of $D_\text{ext} = 20000 \ \text{cm}^2 \ \text{s}^{-1}$. In fact, while a N$^{14}$ excess of 0.53 dex is obtained at the TAMS for a star with $D_\text{ext} = 19000 \ \text{cm}^2 \ \text{s}^{-1}$ and a chemical mixing profile from IGWs (Fig. \ref{fig:DmixDiagram} d), a $D_\text{ext} = 90000 \ \text{cm}^2 \ \text{s}^{-1}$ is needed to get a N$^{14}$ excess of 0.55 dex using a constant radiative envelope mixing (Fig. \ref{fig:DmixDiagram} b).

Figure \ref{fig:Abundance} b shows how the corresponding period spacing patterns at $X_\text{c} = 0.7$, 0.5 and 0.1 varies for the $D_\text{ext}$ range in Fig. \ref{fig:Abundance} a. At $D_\text{ext} \gtrsim 5000 \ \text{cm}^2 \ \text{s}^{-1}$ the overall shape of the patterns no longer changes. Instead they are 'only' shifted towards higher $dP$ values for increasing $D_\text{ext}$. Similarly, the average period spacing is known to increase for increasing stellar mass due to the increase in convective core mass. Therefore, we expand our analyses to check if different combinations of $D_\text{ext}$ and $M_\text{ini}$ will return the same average period spacing of dipole prograde g-modes. To do this we compute another grid of stellar models with exponential diffusive overshooting and a chemical induced radiative envelope mixing profile from IGWs, varying $M_\text{ini}$ from 3.0 to 4.0 M$_\odot$ in steps of 0.05 M$_\odot$ and $D_\text{ext}$ from 1000 to 19000 cm$^2$ s$^{-1}$ in steps of 1000 cm$^2$ s$^{-1}$. For each of the resulting models we calculate average period spacing of dipole prograde g-modes using $dP = \Pi_0/\sqrt{l(l+1)} = \Pi_0/2$. The chemical composition is set to the Galactic standard for B--type stars in the solar neighbourhood and $f_\text{ov} = 0.015$. 

\begin{figure*}
\centering
\includegraphics[width=\linewidth]{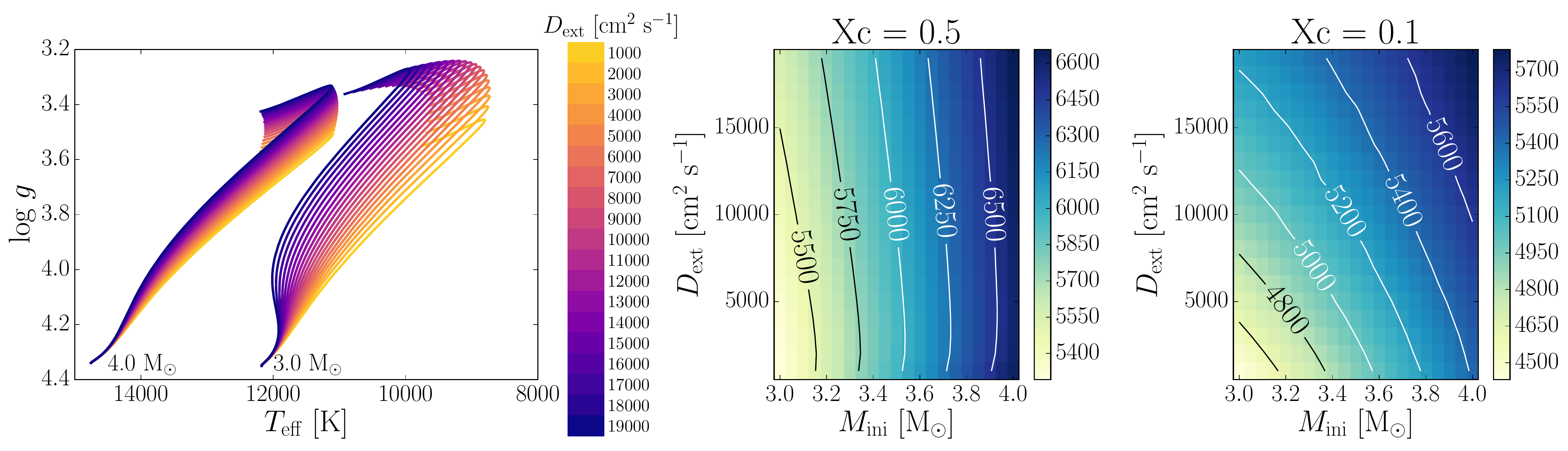}
\caption{\emph{Left panel:} The effect of increasing $D_\text{ext}$ on the evolutionary tracks at two different initial stellar masses, using a $D_\text{mix}$ profile as in Fig. \ref{fig:DmixDiagram} b. The darker the colour of the track, the higher the radiative envelope mixing. \emph{Center and right panels:} 2D surface plots illustrating the correlations between $M_\text{ini}$, $D_\text{ext}$ and $dP$ (given in seconds) at two different $X_\text{c}$. The darker the colour, the higher the average period spacing value of the dipole prograde modes. The contour lines show positions of constant $dP$.}
\label{fig:DextKielCorrelations}
\end{figure*}

The results are displayed in Fig. \ref{fig:DextKielCorrelations}. The left panel illustrates the effect of increasing $D_\text{ext}$ on the evolutionary tracks for two different initial stellar masses, and the center and right panels show correlation plots between $D_\text{ext}$, $M_\text{ini}$ and $dP$ at two different $X_\text{c}$. In the correlation plots in Fig. \ref{fig:DextKielCorrelations} the contour lines illustrate the dependence of $dP$ on $D_\text{ext}$ and $M_\text{ini}$. As seen from the tilt of these lines, the stellar mass is much more important than $D_\text{ext}$ for the obtained average period spacing at $X_\text{c} = 0.5$ and increasing $D_\text{ext}$ only has a small effect. However, this changes towards later stages of the stellar evolution on the MS. 

As the star evolves the contour lines become slanted and a lower stellar mass with a higher envelope mixing results in the same average period spacing for dipole prograde modes as a higher mass star with a lower $D_\text{ext}$. As an example at $X_\text{c} = 0.1$ a star with $M_\text{ini} \sim 3.0 \ \text{M}_\odot$ and $D_\text{ext} \sim 18000 \ \text{cm}^2 \text{s}^{-1}$ has the same average period spacing $dP \sim 5200 \ \text{s}$ as a star with $M_\text{ini} \sim 3.8 \ \text{M}_\odot$ and $D_\text{ext} \sim 1000 \ \text{cm}^2 \text{s}^{-1}$. In other words, if the stellar mass and average period spacing is known for a star with observed N$^{14}$ excess, then it should be possible to distinguish between a constant radiative envelope mixing and a chemical mixing profile from IGWs because of the resulting constraints in the values of $D_\text{ext}$ near the core of the star.

\section{Conclusions}\label{Sec:Conclusions}

We have investigated the capabilities of g-modes pulsations in SPB stars to constrain the shape of convective core overshoot and radiative envelope mixing. Within expected observational frequency errors from the nominal \emph{Kepler} mission, we find that dipole prograde g-modes in the 0.8-3 d period range can be used to distinguish between step and exponential diffusive overshooting. This capability diminishes towards the TAMS at $X_\text{c}= 0.1$ where the g-modes behave the same for both step and exponential diffusive overshooting within the expected frequency errors.

When testing the ability of g-modes to distinguish between different shapes of radiative envelope mixing we find that for dipole prograde g-modes an extended exponential overshooting term and a chemical mixing profile from IGWs results in the same period spacing series as a constant mixing throughout radiative envelope. In other words, the g-modes are not altered by a profile $D_\text{ext}(r)$ since the g-modes only probe the value $D_\text{ext}$ in the near core region. Furthermore, because the extended exponential overshooting has one more free parameter and does not improve the probing power to deduce the core overshoot shape, it is not favourable to use it in forward seismic modelling.

The chemical mixing profile from IGWs is much more efficient than a constant radiative envelope mixing at transporting N$^{14}$ produce through CNO cycle in the core to the surface of the star within the MS lifetime. Combining knowledge from the g-modes together with measured surface N$^{14}$ abundance holds the potential to determine the shape of envelope mixing.

\begin{acknowledgements}
The authors are grateful to Bill Paxton and Richard Townsend and their team of MESA and GYRE developers without whom these results would not have been possible. The authors also thank the referee for providing useful comments and encouragements, as well as some pertinent references to the literature. This research has been funded by the European Research Council (ERC) under the European Union’s Horizon 2020 research and innovation programme (grant agreement N$^\circ$670519: MAMSIE). PIP acknowledges support from The Research Foundation - Flanders (FWO), Belgium.
\end{acknowledgements}



\bibliographystyle{aa}
\bibliography{references.bib}

\begin{thebibliography}{74}
\expandafter\ifx\csname natexlab\endcsname\relax\def\natexlab#1{#1}\fi

\bibitem[{{Aerts} {et~al.}(2010){Aerts}, {Christensen-Dalsgaard}, \&
  {Kurtz}}]{Aerts2010}
{Aerts}, C., {Christensen-Dalsgaard}, J., \& {Kurtz}, D.~W. 2010,
  {Asteroseismology} (Springer, Astronomy and Astrophysics Library), ~ISBN
  978-1-4020-5178-4

\bibitem[{{Aerts} {et~al.}(2014){Aerts}, {Molenberghs}, {Kenward}, \&
  {Neiner}}]{Aerts2014}
{Aerts}, C., {Molenberghs}, G., {Kenward}, M.~G., \& {Neiner}, C. 2014, \apj,
  781, 88

\bibitem[{{Aerts} {et~al.}(2003){Aerts}, {Thoul}, {Daszy{\'n}ska}, {Scuflaire},
  {Waelkens}, {Dupret}, {Niemczura}, \& {Noels}}]{Aerts2003}
{Aerts}, C., {Thoul}, A., {Daszy{\'n}ska}, J., {et~al.} 2003, Science, 300,
  1926

\bibitem[{{Aparicio} {et~al.}(1990){Aparicio}, {Bertelli}, {Chiosi}, \&
  {Garcia-Pelayo}}]{Aparicio1990}
{Aparicio}, A., {Bertelli}, G., {Chiosi}, C., \& {Garcia-Pelayo}, J.~M. 1990,
  \aap, 240, 262

\bibitem[{{Arentoft} {et~al.}(2017){Arentoft}, {Brogaard}, {Jessen-Hansen},
  {Silva Aguirre}, {Kjeldsen}, {Mosumgaard}, \& {Sandquist}}]{Arentoft2017}
{Arentoft}, T., {Brogaard}, K., {Jessen-Hansen}, J., {et~al.} 2017, \apj, 838,
  115

\bibitem[{{Asplund} {et~al.}(2009){Asplund}, {Grevesse}, {Sauval}, \&
  {Scott}}]{Asplund2009}
{Asplund}, M., {Grevesse}, N., {Sauval}, A.~J., \& {Scott}, P. 2009, \araa, 47,
  481

\bibitem[{{Augustson} {et~al.}(2016){Augustson}, {Brun}, \&
  {Toomre}}]{Augustson2016}
{Augustson}, K.~C., {Brun}, A.~S., \& {Toomre}, J. 2016, \apj, 829, 92

\bibitem[{{Auvergne} {et~al.}(2009){Auvergne}, {Bodin}, {Boisnard}, {Buey},
  {Chaintreuil}, {Epstein}, {Jouret}, {Lam-Trong}, {Levacher}, {Magnan},
  {Perez}, {Plasson}, {Plesseria}, {Peter}, {Steller}, {Tiph{\`e}ne}, {Baglin},
  {Agogu{\'e}}, {Appourchaux}, {Barbet}, {Beaufort}, {Bellenger}, {Berlin},
  {Bernardi}, {Blouin}, {Boumier}, {Bonneau}, {Briet}, {Butler}, {Cautain},
  {Chiavassa}, {Costes}, {Cuvilho}, {Cunha-Parro}, {de Oliveira Fialho},
  {Decaudin}, {Defise}, {Djalal}, {Docclo}, {Drummond}, {Dupuis}, {Exil},
  {Faur{\'e}}, {Gaboriaud}, {Gamet}, {Gavalda}, {Grolleau}, {Gueguen},
  {Guivarc'h}, {Guterman}, {Hasiba}, {Huntzinger}, {Hustaix}, {Imbert},
  {Jeanville}, {Johlander}, {Jorda}, {Journoud}, {Karioty}, {Kerjean},
  {Lafond}, {Lapeyrere}, {Landiech}, {Larqu{\'e}}, {Laudet}, {Le Merrer},
  {Leporati}, {Leruyet}, {Levieuge}, {Llebaria}, {Martin}, {Mazy}, {Mesnager},
  {Michel}, {Moalic}, {Monjoin}, {Naudet}, {Neukirchner}, {Nguyen-Kim},
  {Ollivier}, {Orcesi}, {Ottacher}, {Oulali}, {Parisot}, {Perruchot},
  {Piacentino}, {Pinheiro da Silva}, {Platzer}, {Pontet}, {Pradines},
  {Quentin}, {Rohbeck}, {Rolland}, {Rollenhagen}, {Romagnan}, {Russ}, {Samadi},
  {Schmidt}, {Schwartz}, {Sebbag}, {Smit}, {Sunter}, {Tello}, {Toulouse},
  {Ulmer}, {Vandermarcq}, {Vergnault}, {Wallner}, {Waultier}, \&
  {Zanatta}}]{Auvergne2009}
{Auvergne}, M., {Bodin}, P., {Boisnard}, L., {et~al.} 2009, \aap, 506, 411

\bibitem[{{Bailey} {et~al.}(2015){Bailey}, {Nagayama}, {Loisel}, {Rochau},
  {Blancard}, {Colgan}, {Cosse}, {Faussurier}, {Fontes}, {Gilleron},
  {Golovkin}, {Hansen}, {Iglesias}, {Kilcrease}, {MacFarlane}, {Mancini},
  {Nahar}, {Orban}, {Pain}, {Pradhan}, {Sherrill}, \& {Wilson}}]{Bailey2015}
{Bailey}, J.~E., {Nagayama}, T., {Loisel}, G.~P., {et~al.} 2015, Nature, 517,
  56

\bibitem[{{Battino} {et~al.}(2016){Battino}, {Pignatari}, {Ritter}, {Herwig},
  {Denisenkov}, {Den Hartogh}, {Trappitsch}, {Hirschi}, {Freytag},
  {Thielemann}, \& {Paxton}}]{Battino2016}
{Battino}, U., {Pignatari}, M., {Ritter}, C., {et~al.} 2016, \apj, 827, 30

\bibitem[{{B{\"o}hm-Vitense}(1958)}]{BohnVitense1958}
{B{\"o}hm-Vitense}, E. 1958, \zap, 46, 108

\bibitem[{{Borucki} {et~al.}(2010){Borucki}, {Koch}, {Basri}, {Batalha},
  {Brown}, {Caldwell}, {Caldwell}, {Christensen-Dalsgaard}, {Cochran},
  {DeVore}, {Dunham}, {Dupree}, {Gautier}, {Geary}, {Gilliland}, {Gould},
  {Howell}, {Jenkins}, {Kondo}, {Latham}, {Marcy}, {Meibom}, {Kjeldsen},
  {Lissauer}, {Monet}, {Morrison}, {Sasselov}, {Tarter}, {Boss}, {Brownlee},
  {Owen}, {Buzasi}, {Charbonneau}, {Doyle}, {Fortney}, {Ford}, {Holman},
  {Seager}, {Steffen}, {Welsh}, {Rowe}, {Anderson}, {Buchhave}, {Ciardi},
  {Walkowicz}, {Sherry}, {Horch}, {Isaacson}, {Everett}, {Fischer}, {Torres},
  {Johnson}, {Endl}, {MacQueen}, {Bryson}, {Dotson}, {Haas}, {Kolodziejczak},
  {Van Cleve}, {Chandrasekaran}, {Twicken}, {Quintana}, {Clarke}, {Allen},
  {Li}, {Wu}, {Tenenbaum}, {Verner}, {Bruhweiler}, {Barnes}, \&
  {Prsa}}]{Borucki2010}
{Borucki}, W.~J., {Koch}, D., {Basri}, G., {et~al.} 2010, Science, 327, 977

\bibitem[{{Brott} {et~al.}(2011){Brott}, {Evans}, {Hunter}, {de Koter},
  {Langer}, {Dufton}, {Cantiello}, {Trundle}, {Lennon}, {de Mink}, {Yoon}, \&
  {Anders}}]{Brott2011}
{Brott}, I., {Evans}, C.~J., {Hunter}, I., {et~al.} 2011, \aap, 530, A116

\bibitem[{{Browning} {et~al.}(2004){Browning}, {Brun}, \&
  {Toomre}}]{Browning2004}
{Browning}, M.~K., {Brun}, A.~S., \& {Toomre}, J. 2004, \apj, 601, 512

\bibitem[{{Castelli} \& {Kurucz}(2003)}]{Castelli2003}
{Castelli}, F. \& {Kurucz}, R.~L. 2003, in IAU Symposium, Vol. 210, Modelling
  of Stellar Atmospheres, ed. N.~{Piskunov}, W.~W. {Weiss}, \& D.~F. {Gray},
  A20

\bibitem[{{Claret} \& {Torres}(2016)}]{Claret2016}
{Claret}, A. \& {Torres}, G. 2016, \aap, 592, A15

\bibitem[{{Claret} \& {Torres}(2017)}]{Claret2017}
{Claret}, A. \& {Torres}, G. 2017, \apj, 849, 18

\bibitem[{{Cox} \& {Giuli}(1968)}]{Cox1968}
{Cox}, J.~P. \& {Giuli}, R.~T. 1968, {Principles of stellar structure }

\bibitem[{{Degroote} {et~al.}(2010){Degroote}, {Aerts}, {Baglin}, {Miglio},
  {Briquet}, {Noels}, {Niemczura}, {Montalban}, {Bloemen}, {Oreiro}, {Vu{\v
  c}kovi{\'c}}, {Smolders}, {Auvergne}, {Baudin}, {Catala}, \&
  {Michel}}]{Degroote2010}
{Degroote}, P., {Aerts}, C., {Baglin}, A., {et~al.} 2010, \nat, 464, 259

\bibitem[{{Deheuvels} {et~al.}(2016){Deheuvels}, {Brand{\~a}o}, {Silva
  Aguirre}, {Ballot}, {Michel}, {Cunha}, {Lebreton}, \&
  {Appourchaux}}]{Deheuvels2016}
{Deheuvels}, S., {Brand{\~a}o}, I., {Silva Aguirre}, V., {et~al.} 2016, \aap,
  589, A93

\bibitem[{{Deheuvels} \& {Michel}(2011)}]{Deheuvels2011}
{Deheuvels}, S. \& {Michel}, E. 2011, \aap, 535, A91

\bibitem[{{Fadeyev}(2015)}]{Fadeyev2015}
{Fadeyev}, Y.~A. 2015, \mnras, 449, 1011

\bibitem[{{Freytag} {et~al.}(1996){Freytag}, {Ludwig}, \&
  {Steffen}}]{Freytag1996}
{Freytag}, B., {Ludwig}, H.-G., \& {Steffen}, M. 1996, \aap, 313, 497

\bibitem[{{Gilet} {et~al.}(2013){Gilet}, {Almgren}, {Bell}, {Nonaka},
  {Woosley}, \& {Zingale}}]{Gilet2013}
{Gilet}, C., {Almgren}, A.~S., {Bell}, J.~B., {et~al.} 2013, \apj, 773, 137

\bibitem[{{Groenewegen} {et~al.}(2007){Groenewegen}, {Decin}, {Salaris}, \& {De
  Cat}}]{Groenewegen2007}
{Groenewegen}, M.~A.~T., {Decin}, L., {Salaris}, M., \& {De Cat}, P. 2007,
  \aap, 463, 579

\bibitem[{{Guinan} {et~al.}(2000){Guinan}, {Ribas}, {Fitzpatrick},
  {Gim{\'e}nez}, {Jordi}, {McCook}, \& {Popper}}]{Guinan2000}
{Guinan}, E.~F., {Ribas}, I., {Fitzpatrick}, E.~L., {et~al.} 2000, \apj, 544,
  409

\bibitem[{{Hauschildt} {et~al.}(1999{\natexlab{a}}){Hauschildt}, {Allard}, \&
  {Baron}}]{Hauschildt1999a}
{Hauschildt}, P.~H., {Allard}, F., \& {Baron}, E. 1999{\natexlab{a}}, \apj,
  512, 377

\bibitem[{{Hauschildt} {et~al.}(1999{\natexlab{b}}){Hauschildt}, {Allard},
  {Ferguson}, {Baron}, \& {Alexander}}]{Hauschildt1999b}
{Hauschildt}, P.~H., {Allard}, F., {Ferguson}, J., {Baron}, E., \& {Alexander},
  D.~R. 1999{\natexlab{b}}, \apj, 525, 871

\bibitem[{{Herwig}(2000)}]{Herwig2000}
{Herwig}, F. 2000, \aap, 360, 952

\bibitem[{{Herwig} {et~al.}(2007){Herwig}, {Freytag}, {Fuchs}, {Hansen},
  {Hueckstaedt}, {Porter}, {Timmes}, \& {Woodward}}]{Herwig2007}
{Herwig}, F., {Freytag}, B., {Fuchs}, T., {et~al.} 2007, in Astronomical
  Society of the Pacific Conference Series, Vol. 378, Why Galaxies Care About
  AGB Stars: Their Importance as Actors and Probes, ed. F.~{Kerschbaum},
  C.~{Charbonnel}, \& R.~F. {Wing}, 43

\bibitem[{{Hj{\o}rringgaard} {et~al.}(2017){Hj{\o}rringgaard}, {Silva Aguirre},
  {White}, {Huber}, {Pope}, {Casagrande}, {Justesen}, \&
  {Christensen-Dalsgaard}}]{Hjorringgaard2017}
{Hj{\o}rringgaard}, J.~G., {Silva Aguirre}, V., {White}, T.~R., {et~al.} 2017,
  \mnras, 464, 3713

\bibitem[{{Hunter} {et~al.}(2008){Hunter}, {Brott}, {Lennon}, {Langer},
  {Dufton}, {Trundle}, {Smartt}, {de Koter}, {Evans}, \& {Ryans}}]{Hunter2008}
{Hunter}, I., {Brott}, I., {Lennon}, D.~J., {et~al.} 2008, \apjl, 676, L29

\bibitem[{{Kozhurina-Platais} {et~al.}(1997){Kozhurina-Platais}, {Demarque},
  {Platais}, {Orosz}, \& {Barnes}}]{Kozhurina-Platais1997}
{Kozhurina-Platais}, V., {Demarque}, P., {Platais}, I., {Orosz}, J.~A., \&
  {Barnes}, S. 1997, \aj, 113, 1045

\bibitem[{{Lacy} {et~al.}(2012){Lacy}, {Torres}, {Fekel}, {Sabby}, \&
  {Claret}}]{Lacy2012}
{Lacy}, C.~H.~S., {Torres}, G., {Fekel}, F.~C., {Sabby}, J.~A., \& {Claret}, A.
  2012, \aj, 143, 129

\bibitem[{{Lenz} {et~al.}(2010){Lenz}, {Pamyatnykh}, {Zdravkov}, \&
  {Breger}}]{Lenz2010}
{Lenz}, P., {Pamyatnykh}, A.~A., {Zdravkov}, T., \& {Breger}, M. 2010, \aap,
  509, A90

\bibitem[{{Liu} {et~al.}(2014){Liu}, {Yang}, {Bi}, {Tian}, {Liu}, {Ge}, {Yu},
  {Li}, {Tan}, {He}, {Wu}, \& {Chintarungruangchai}}]{Liu2014}
{Liu}, Z., {Yang}, W., {Bi}, S., {et~al.} 2014, \apj, 780, 152

\bibitem[{{Maeder} \& {Mermilliod}(1981)}]{Maeder1981}
{Maeder}, A. \& {Mermilliod}, J.~C. 1981, \aap, 93, 136

\bibitem[{{Martins} {et~al.}(2015){Martins}, {Sim{\'o}n-D{\'{\i}}az},
  {Palacios}, {Howarth}, {Georgy}, {Walborn}, {Bouret}, \&
  {Barb{\'a}}}]{Martins2015}
{Martins}, F., {Sim{\'o}n-D{\'{\i}}az}, S., {Palacios}, A., {et~al.} 2015,
  \aap, 578, A109

\bibitem[{{Mathis}(2013)}]{Mathis2013}
{Mathis}, S. 2013, in Lecture Notes in Physics, Berlin Springer Verlag, Vol.
  865, Lecture Notes in Physics, Berlin Springer Verlag, ed. M.~{Goupil},
  K.~{Belkacem}, C.~{Neiner}, F.~{Ligni{\`e}res}, \& J.~J. {Green}, 23

\bibitem[{{Mathis} {et~al.}(2004){Mathis}, {Palacios}, \& {Zahn}}]{Mathis2004}
{Mathis}, S., {Palacios}, A., \& {Zahn}, J.-P. 2004, \aap, 425, 243

\bibitem[{{Meakin} \& {Arnett}(2007)}]{Meakin2007}
{Meakin}, C.~A. \& {Arnett}, D. 2007, \apj, 667, 448

\bibitem[{{Meynet} {et~al.}(1993){Meynet}, {Mermilliod}, \&
  {Maeder}}]{Meynet1993}
{Meynet}, G., {Mermilliod}, J.-C., \& {Maeder}, A. 1993, \aaps, 98, 477

\bibitem[{{Miglio} {et~al.}(2008){Miglio}, {Montalb{\'a}n}, {Noels}, \&
  {Eggenberger}}]{Miglio2008}
{Miglio}, A., {Montalb{\'a}n}, J., {Noels}, A., \& {Eggenberger}, P. 2008,
  \mnras, 386, 1487

\bibitem[{{Montalb{\'a}n} {et~al.}(2013){Montalb{\'a}n}, {Miglio}, {Noels},
  {Dupret}, {Scuflaire}, \& {Ventura}}]{Montalban2013}
{Montalb{\'a}n}, J., {Miglio}, A., {Noels}, A., {et~al.} 2013, \apj, 766, 118

\bibitem[{{Moravveji}(2016)}]{Moravveji2016MNRAS.455L..67M}
{Moravveji}, E. 2016, \mnras, 455, L67

\bibitem[{{Moravveji} {et~al.}(2015){Moravveji}, {Aerts}, {P{\'a}pics},
  {Triana}, \& {Vandoren}}]{Moravveji2015}
{Moravveji}, E., {Aerts}, C., {P{\'a}pics}, P.~I., {Triana}, S.~A., \&
  {Vandoren}, B. 2015, \aap, 580, A27

\bibitem[{{Moravveji} {et~al.}(2016){Moravveji}, {Townsend}, {Aerts}, \&
  {Mathis}}]{Moravveji2016ApJ...823..130M}
{Moravveji}, E., {Townsend}, R.~H.~D., {Aerts}, C., \& {Mathis}, S. 2016, \apj,
  823, 130

\bibitem[{{Neiner} {et~al.}(2012){Neiner}, {Mathis}, {Saio}, {Lovekin},
  {Eggenberger}, \& {Lee}}]{Neiner2012}
{Neiner}, C., {Mathis}, S., {Saio}, H., {et~al.} 2012, \aap, 539, A90

\bibitem[{{Nieva} \& {Przybilla}(2012)}]{Nieva2012}
{Nieva}, M.-F. \& {Przybilla}, N. 2012, \aap, 539, A143

\bibitem[{{Pamyatnykh} {et~al.}(2004){Pamyatnykh}, {Handler}, \&
  {Dziembowski}}]{Pamyatnykh2004}
{Pamyatnykh}, A.~A., {Handler}, G., \& {Dziembowski}, W.~A. 2004, \mnras, 350,
  1022

\bibitem[{{P{\'a}pics} {et~al.}(2014){P{\'a}pics}, {Moravveji}, {Aerts},
  {Tkachenko}, {Triana}, {Bloemen}, \& {Southworth}}]{Papics2014}
{P{\'a}pics}, P.~I., {Moravveji}, E., {Aerts}, C., {et~al.} 2014, \aap, 570, A8

\bibitem[{{P{\'a}pics} {et~al.}(2015){P{\'a}pics}, {Tkachenko}, {Aerts}, {Van
  Reeth}, {De Smedt}, {Hillen}, {{\O}stensen}, \& {Moravveji}}]{Papics2015}
{P{\'a}pics}, P.~I., {Tkachenko}, A., {Aerts}, C., {et~al.} 2015, \apjl, 803,
  L25

\bibitem[{{P{\'a}pics} {et~al.}(2017){P{\'a}pics}, {Tkachenko}, {Van Reeth},
  {Aerts}, {Moravveji}, {Van de Sande}, {De Smedt}, {Bloemen}, {Southworth},
  {Debosscher}, {Niemczura}, \& {Gameiro}}]{Papics2017}
{P{\'a}pics}, P.~I., {Tkachenko}, A., {Van Reeth}, T., {et~al.} 2017, \aap,
  598, A74

\bibitem[{{Paxton} {et~al.}(2011){Paxton}, {Bildsten}, {Dotter}, {Herwig},
  {Lesaffre}, \& {Timmes}}]{Paxton2011}
{Paxton}, B., {Bildsten}, L., {Dotter}, A., {et~al.} 2011, \apjs, 192, 3

\bibitem[{{Paxton} {et~al.}(2013){Paxton}, {Cantiello}, {Arras}, {Bildsten},
  {Brown}, {Dotter}, {Mankovich}, {Montgomery}, {Stello}, {Timmes}, \&
  {Townsend}}]{Paxton2013}
{Paxton}, B., {Cantiello}, M., {Arras}, P., {et~al.} 2013, \apjs, 208, 4

\bibitem[{{Paxton} {et~al.}(2015){Paxton}, {Marchant}, {Schwab}, {Bauer},
  {Bildsten}, {Cantiello}, {Dessart}, {Farmer}, {Hu}, {Langer}, {Townsend},
  {Townsley}, \& {Timmes}}]{Paxton2015}
{Paxton}, B., {Marchant}, P., {Schwab}, J., {et~al.} 2015, \apjs, 220, 15

\bibitem[{{Prada Moroni} {et~al.}(2012){Prada Moroni}, {Gennaro}, {Bono},
  {Pietrzy{\'n}ski}, {Gieren}, {Pilecki}, {Graczyk}, \& {Thompson}}]{Prada2012}
{Prada Moroni}, P.~G., {Gennaro}, M., {Bono}, G., {et~al.} 2012, \apj, 749, 108

\bibitem[{{Prat} {et~al.}(2016){Prat}, {Guilet}, {Viallet}, \&
  {M{\"u}ller}}]{Prat2016}
{Prat}, V., {Guilet}, J., {Viallet}, M., \& {M{\"u}ller}, E. 2016, \aap, 592,
  A59

\bibitem[{{Przybilla} {et~al.}(2013){Przybilla}, {Nieva}, {Irrgang}, \&
  {Butler}}]{Przybilla2013}
{Przybilla}, N., {Nieva}, M.~F., {Irrgang}, A., \& {Butler}, K. 2013, in EAS
  Publications Series, Vol.~63, EAS Publications Series, ed. G.~{Alecian},
  Y.~{Lebreton}, O.~{Richard}, \& G.~{Vauclair}, 13--23

\bibitem[{{Ribas} {et~al.}(2000){Ribas}, {Jordi}, \& {Gim{\'e}nez}}]{Ribas2000}
{Ribas}, I., {Jordi}, C., \& {Gim{\'e}nez}, {\'A}. 2000, \mnras, 318, L55

\bibitem[{{Rogers} {et~al.}(2013){Rogers}, {Lin}, {McElwaine}, \&
  {Lau}}]{Rogers2013}
{Rogers}, T.~M., {Lin}, D.~N.~C., {McElwaine}, J.~N., \& {Lau}, H.~H.~B. 2013,
  \apj, 772, 21

\bibitem[{{Rogers} \& {McElwaine}(2017)}]{Rogers2017}
{Rogers}, T.~M. \& {McElwaine}, J.~N. 2017, \apjl, 848, L1

\bibitem[{{Rosenfield} {et~al.}(2017){Rosenfield}, {Girardi}, {Williams},
  {Johnson}, {Dolphin}, {Bressan}, {Weisz}, {Dalcanton}, {Fouesneau}, \&
  {Kalirai}}]{Rosenfield2017}
{Rosenfield}, P., {Girardi}, L., {Williams}, B.~F., {et~al.} 2017, \apj, 841,
  69

\bibitem[{{Schmid} \& {Aerts}(2016)}]{Schmid2016}
{Schmid}, V.~S. \& {Aerts}, C. 2016, \aap, 592, A116

\bibitem[{{Silva Aguirre} {et~al.}(2011){Silva Aguirre}, {Ballot}, {Serenelli},
  \& {Weiss}}]{SilvaAguirre2011}
{Silva Aguirre}, V., {Ballot}, J., {Serenelli}, A.~M., \& {Weiss}, A. 2011,
  \aap, 529, A63

\bibitem[{{Stancliffe} {et~al.}(2015){Stancliffe}, {Fossati}, {Passy}, \&
  {Schneider}}]{Stancliffe2015}
{Stancliffe}, R.~J., {Fossati}, L., {Passy}, J.-C., \& {Schneider}, F.~R.~N.
  2015, \aap, 575, A117

\bibitem[{{Townsend} \& {Teitler}(2013)}]{Townsend2013}
{Townsend}, R.~H.~D. \& {Teitler}, S.~A. 2013, \mnras, 435, 3406

\bibitem[{{Triana} {et~al.}(2015){Triana}, {Moravveji}, {P{\'a}pics}, {Aerts},
  {Kawaler}, \& {Christensen-Dalsgaard}}]{Triana2015}
{Triana}, S.~A., {Moravveji}, E., {P{\'a}pics}, P.~I., {et~al.} 2015, \apj,
  810, 16

\bibitem[{{Van Reeth} {et~al.}(2015){Van Reeth}, {Tkachenko}, {Aerts},
  {P{\'a}pics}, {Triana}, {Zwintz}, {Degroote}, {Debosscher}, {Bloemen},
  {Schmid}, {De Smedt}, {Fremat}, {Fuentes}, {Homan}, {Hrudkova},
  {Karjalainen}, {Lombaert}, {Nemeth}, {{\O}stensen}, {Van De Steene}, {Vos},
  {Raskin}, \& {Van Winckel}}]{VanReeth2015}
{Van Reeth}, T., {Tkachenko}, A., {Aerts}, C., {et~al.} 2015, \apjs, 218, 27

\bibitem[{{VandenBerg} \& {Stetson}(2004)}]{VandenBerg2004}
{VandenBerg}, D.~A. \& {Stetson}, P.~B. 2004, \pasp, 116, 997

\bibitem[{{Viallet} {et~al.}(2015){Viallet}, {Meakin}, {Prat}, \&
  {Arnett}}]{Viallet2015}
{Viallet}, M., {Meakin}, C., {Prat}, V., \& {Arnett}, D. 2015, \aap, 580, A61

\bibitem[{{Walczak} \& {Handler}(2015)}]{Walczak2015}
{Walczak}, P. \& {Handler}, G. 2015, in IAU Symposium, Vol. 307, New Windows on
  Massive Stars, ed. G.~{Meynet}, C.~{Georgy}, J.~{Groh}, \& P.~{Stee},
  239--240

\bibitem[{{Yang} {et~al.}(2015){Yang}, {Tian}, {Bi}, {Ge}, {Wu}, \&
  {Zhang}}]{Yang2015}
{Yang}, W., {Tian}, Z., {Bi}, S., {et~al.} 2015, \mnras, 453, 2094

\bibitem[{{Zahn}(2011)}]{Zahn2011}
{Zahn}, J.-P. 2011, in IAU Symposium, Vol. 272, Active OB Stars: Structure,
  Evolution, Mass Loss, and Critical Limits, ed. C.~{Neiner}, G.~{Wade},
  G.~{Meynet}, \& G.~{Peters}, 14--25

\end{thebibliography}


\appendix

\section{MESA inlist file}\label{App:MESAinlists}

The MESA input parameters are given in the form of an inlist file, in which all parameters that the user wants to change from the default settings are specified. Below is the inlist setup used to compute the stellar models in this work. Parameters that have not been given a value in this list, i.e. have nothing following the equality sign, are the ones that we vary. The varied parameters and MESA input parameters are related as follows: \textsf{new\_Y =} $1 - 0.014 - X_\text{ini}$, \textsf{initial\_mass =} $M_\text{ini}$, \textsf{step\_overshoot\_f\_above\_burn\_h\_core =} $\alpha_\text{ov}$, \textsf{overshoot\_f\_above\_burn\_h\_core =} $f_\text{ov}$ and $f_\text{1}$, \textsf{overshoot\_D2\_above\_burn\_h\_core =} $D_\text{2}$, \textsf{overshoot\_f2\_above\_burn\_h\_core =} $f_\text{2}$ and \textsf{min\_D\_mix =} $D_\text{ext}$. The MESA inlist setup is:\\

\setlength{\leftskip}{.1cm}
\textsf{\&star\_job}\\

\setlength{\leftskip}{0.5cm}
\indent \textsf{show\_log\_description\_at\_start = .false.}\\
\indent \textsf{show\_net\_species\_info = .false.}\\

\indent \textsf{create\_pre\_main\_sequence\_model = .false.}\\
\indent \textsf{pgstar\_flag = .false.}\\

\indent \textsf{change\_lnPgas\_flag = .true.}\\
\indent \textsf{change\_initial\_lnPgas\_flag = .true.}\\
\indent \textsf{new\_lnPgas\_flag = .true.}\\

\indent \textsf{change\_net = .true.}\\
\indent \textsf{new\_net\_name = 'pp\_cno\_extras\_o18\_ne22.net'}\\
\indent \textsf{change\_initial\_net = .true.}\\
\indent \textsf{auto\_extend\_net = .true.}\\

\indent \textsf{initial\_zfracs = 6    ! Asplund et al. (2009)}\\

\indent \textsf{kappa\_blend\_logT\_upper\_bdy = 4.5d0}\\
\indent \textsf{kappa\_blend\_logT\_lower\_bdy = 4.5d0}\\
\indent \textsf{kappa\_lowT\_prefix = 'lowT\_fa05\_a09p'}\\

\indent \textsf{kappa\_file\_prefix = 'Mono\_a09\_Fe1.75\_Ni1.75'}\\
\indent \textsf{kappa\_CO\_prefix = 'a09\_co'}\\

\indent \textsf{relax\_Y = .true.}\\
\indent \textsf{change\_Y = .true.}\\
\indent \textsf{relax\_initial\_Y = .true.}\\
\indent \textsf{change\_initial\_Y = .true.}\\
\indent \textsf{new\_Y = }\\

\indent \textsf{relax\_Z = .true.}\\
\indent \textsf{change\_Z = .true.}\\
\indent \textsf{relax\_initial\_Z = .true.}\\
\indent \textsf{change\_initial\_Z = .true.}\\
\indent \textsf{new\_Z = 0.014}\\

\setlength{\leftskip}{.1cm}
\indent \textsf{/ !end of star\_job namelist}\\

\textsf{\&controls}\\

\setlength{\leftskip}{0.5cm}
\indent \textsf{initial\_mass = }\\
\indent \textsf{log\_directory = }\\

\indent \textsf{mixing\_length\_alpha = 2.0}\\

\indent \textsf{set\_min\_D\_mix = .true.}\\
\indent \textsf{min\_D\_mix = }\\

\indent \textsf{overshoot\_f0\_above\_burn\_h\_core = 0.001}\\
\indent \textsf{step\_overshoot\_f\_above\_burn\_h\_core = }\\
\indent \textsf{overshoot\_f\_above\_burn\_h\_core = }\\
\indent \textsf{overshoot\_D2\_above\_burn\_h = }\\
\indent \textsf{overshoot\_f2\_above\_burn\_h = }\\

\indent \textsf{max\_years\_for\_timestep = 1.0d5}\\
\indent \textsf{varcontrol\_target = 5d-5}\\

\indent \textsf{delta\_lg\_XH\_cntr\_max = -1}\\
\indent \textsf{delta\_lg\_XH\_cntr\_limit = 0.05}\\

\indent \textsf{alpha\_semiconvection = 0.01}\\

\indent \textsf{write\_pulse\_info\_with\_profile = .true.}\\
\indent \textsf{pulse\_info\_format = 'GYRE'}\\

\indent \textsf{xa\_central\_lower\_limit\_species(1) = 'h1'}\\
\indent \textsf{xa\_central\_lower\_limit(1) = 1d-3}\\
\indent \textsf{when\_to\_stop\_rtol = 1d-3}\\
\indent \textsf{when\_to\_stop\_atol = 1d-3}\\

\indent \textsf{terminal\_interval = 25}\\
\indent \textsf{write\_header\_frequency = 4}\\
\indent \textsf{photostep = 500}\\
\indent \textsf{history\_interval = 1}\\
\indent \textsf{write\_profiles\_flag = .false.}\\
\indent \textsf{mixing\_D\_limit\_for\_log = 1d-4}\\

\indent \textsf{use\_Ledoux\_criterion = .true.}\\
\indent \textsf{num\_cells\_for\_smooth\_gradL\_composition\_term = 0}\\
\indent \textsf{D\_mix\_ov\_limit = 0d0}\\

\indent \textsf{which\_atm\_option = 'photosphere\_tables'}\\

\indent \textsf{calculate\_Brunt\_N2 = .true.}\\
\indent \textsf{num\_cells\_for\_smooth\_brunt\_B = 0}\\

\indent \textsf{cubic\_interpolation\_in\_Z = .true.}\\
\indent \textsf{use\_Type2\_opacities = .false.}\\
\indent \textsf{kap\_Type2\_full\_off\_X = 1d-6}\\
\indent \textsf{kap\_Type2\_full\_on\_X = 1d-6}\\

\indent \textsf{! Uncomment the following line when using the implemented radiative envelope mixing profile from \citet{Rogers2017}}\\
\indent \textsf{!use\_other\_D\_mix = .true.}\\

\indent \textsf{mesh\_delta\_coeff = 0.2}\\
\indent \textsf{max\_allowed\_nz = 35000}\\
\indent \textsf{max\_dq = 1d-3}\\

\indent \textsf{R\_function\_weight = 10}\\
\indent \textsf{R\_function2\_weight = 10}\\
\indent \textsf{R\_function2\_param1 = 1000}\\

\indent \textsf{xtra\_coef\_above\_xtrans = 0.2}\\
\indent \textsf{xtra\_coef\_below\_xtrans = 0.2}\\
\indent \textsf{xtra\_dist\_above\_xtrans = 0.5}\\
\indent \textsf{xtra\_dist\_below\_xtrans = 0.5}\\

\indent \textsf{mesh\_logX\_species(1) = 'h1'}\\
\indent \textsf{mesh\_logX\_min\_for\_extra(1) = -12}\\
\indent \textsf{mesh\_dlogX\_dlogP\_extra(1) = 0.15}\\
\indent \textsf{mesh\_dlogX\_dlogP\_full\_on(1) = 1d-6}\\
\indent \textsf{mesh\_dlogX\_dlogP\_full\_off(1) = 1d-12}\\

\indent \textsf{mesh\_logX\_species(2) = 'he4'}\\
\indent \textsf{mesh\_logX\_min\_for\_extra(2) = -12}\\
\indent \textsf{mesh\_dlogX\_dlogP\_extra(2) = 0.15}\\
\indent \textsf{mesh\_dlogX\_dlogP\_full\_on(2) = 1d-6}\\
\indent \textsf{mesh\_dlogX\_dlogP\_full\_off(2) = 1d-12}\\

\indent \textsf{mesh\_logX\_species(3) = 'n14'}\\
\indent \textsf{mesh\_logX\_min\_for\_extra(3) = -12}\\
\indent \textsf{mesh\_dlogX\_dlogP\_extra(3) = 0.15}\\
\indent \textsf{mesh\_dlogX\_dlogP\_full\_on(3) = 1d-6}\\
\indent \textsf{mesh\_dlogX\_dlogP\_full\_off(3) = 1d-12}\\

\indent \textsf{P\_function\_weight = 30}\\
\indent \textsf{T\_function1\_weight = 75}\\

\indent \textsf{xa\_function\_species(1) = 'h1'}\\
\indent \textsf{xa\_function\_weight(1) = 80}\\
\indent \textsf{xa\_function\_param(1) = 1d-2}\\

\indent \textsf{xa\_function\_species(2) = 'he4'}\\
\indent \textsf{xa\_function\_weight(2) = 80}\\
\indent \textsf{xa\_function\_param(2) = 1d-2}\\

\setlength{\leftskip}{.1cm}
\indent \textsf{/ ! end of controls namelist}\\

\setlength{\leftskip}{0cm}

\section{GYRE inlist file}\label{App:GYREinlist}

The following displays the setup of the GYRE inlist file used to compute the pulsation properties for the stellar models used in this paper. Parameters that have not been filled in this list are varied parameters as well as input and output file names. \\

\textsf{\&constants}\\
\setlength{\leftskip}{.1cm}
\indent \textsf{/}\\

\textsf{\&model}\\
\setlength{\leftskip}{0.5cm}
\indent \indent \textsf{model\_type = 'EVOL'}\\
\indent \indent \textsf{file = ' '}\\
\indent \indent \textsf{file\_format = 'MESA'}\\
\indent \indent \textsf{reconstruct\_As = .False.}\\
\indent \indent \textsf{uniform\_rotation= .True.}\\
\indent \indent \textsf{Omega\_uni= 0.0}\\
\setlength{\leftskip}{.1cm}
\indent \textsf{/}\\

\textsf{\&osc}\\
\setlength{\leftskip}{0.5cm}
\indent \indent \textsf{outer\_bound = 'ZERO'}\\
\indent \indent \textsf{rotation\_method = 'TRAD'}\\
\setlength{\leftskip}{.1cm}
\indent \textsf{/}\\

\textsf{\&mode}\\
\setlength{\leftskip}{0.5cm}
\indent \indent \textsf{l = 1}\\
\indent \indent \textsf{m = 1}\\
\indent \indent \textsf{n\_pg\_min = -75}\\
\indent \indent \textsf{n\_pg\_max = -5}\\
\setlength{\leftskip}{.1cm}
\indent \textsf{/}\\

\textsf{\&num}\\
\setlength{\leftskip}{0.5cm}
\indent \indent \textsf{ivp\_solver = 'MAGNUS\_GL4'}\\
\setlength{\leftskip}{.1cm}
\indent \textsf{/}\\

\textsf{\&scan}\\
\setlength{\leftskip}{0.5cm}
\indent \indent \textsf{grid\_type = 'INVERSE'}\\
\indent \indent \textsf{grid\_frame = 'COROT\_I'}\\
\indent \indent \textsf{freq\_units = 'PER\_DAY'}\\
\indent \indent \textsf{freq\_frame = 'INERTIAL'}\\
\indent \indent \textsf{freq\_min = }\\
\indent \indent \textsf{freq\_max = }\\
\indent \indent \textsf{n\_freq = 400}\\
\setlength{\leftskip}{.1cm}
\indent \textsf{/}\\

\textsf{\&shoot\_grid}\\
\setlength{\leftskip}{0.5cm}
\indent \indent \textsf{op\_type = 'CREATE\_CLONE'}\\
\setlength{\leftskip}{.1cm}
\indent \textsf{/}\\

\textsf{\&recon\_grid}\\
\setlength{\leftskip}{0.5cm}
\indent \indent \textsf{op\_type = 'CREATE\_CLONE'}\\
\setlength{\leftskip}{.1cm}
\indent \textsf{/}\\

\textsf{\&shoot\_grid}\\
\setlength{\leftskip}{0.5cm}
\indent \indent \textsf{op\_type = 'RESAMP\_CENTER'}\\
\indent \indent \textsf{n = 12}\\
\setlength{\leftskip}{.1cm}
\indent \textsf{/}\\

\textsf{\&shoot\_grid}\\
\setlength{\leftskip}{0.5cm}
\indent \indent \textsf{op\_type = 'RESAMP\_DISPERSION'}\\
\indent \indent \textsf{alpha\_osc = 5}\\
\indent \indent \textsf{alpha\_exp = 1}\\
\setlength{\leftskip}{.1cm}
\indent \textsf{/}\\

\textsf{\&recon\_grid}\\
\setlength{\leftskip}{0.5cm}
\indent \indent \textsf{op\_type = 'RESAMP\_CENTER'}\\
\indent \indent \textsf{n = 12}\\
\setlength{\leftskip}{.1cm}
\indent \textsf{/}\\

\textsf{\&recon\_grid}\\
\setlength{\leftskip}{0.5cm}
\indent \indent \textsf{op\_type = 'RESAMP\_DISPERSION'}\\
\indent \indent \textsf{alpha\_osc = 5}\\
\indent \indent \textsf{alpha\_exp = 1}\\
\setlength{\leftskip}{.1cm}
\indent \textsf{/}\\

\textsf{\&output}\\
\setlength{\leftskip}{0.5cm}
\indent \indent \textsf{summary\_file = ' '}\\
\indent \indent \textsf{summary\_file\_format = 'TXT'}\\
\indent \indent \textsf{summary\_item\_list = 'M\_star, R\_star, beta, l, n\_pg, omega, freq, freq\_units, E\_norm'}\\
\indent \indent \textsf{mode\_prefix = ''}\\
\indent \indent \textsf{mode\_file\_format = 'HDF'}\\
\indent \indent \textsf{mode\_item\_list = 'l, beta, n\_pg, omega, freq, freq\_units, x, xi\_r, xi\_h, K'}\\
\indent \indent \textsf{freq\_units = 'PER\_DAY'}\\
\setlength{\leftskip}{.1cm}
\indent \textsf{/}\\

\section{Internal grid comparisons}\label{App:Grid}

We investigate if any set of parameters within the exponential and step overshooting grids essentially returns the same best matching models as according to the merit function cut-off defined in Sect. \ref{Sec:MF}. In other words, we compare benchmark models A and B (C and D) against the exponential diffusive (step) overshooting grids specified in Table \ref{Tab:GridParameters}. Not surprisingly, the best matching model return the exact same model parameters as for benchmark model A/C and B/D, resulting in MF$_\text{Best} = 0$ in all four cases. The merit function cut-offs below which the models match the benchmark models equally well within expected observational frequency errors of \textit{Kepler} data are MF$_\text{cut,A} = 1.41$ and $\text{MF}_\text{cut,C} = 1.42$ for benchmark model A (C) compared to the exponential (step) overshooting grid, MF$_\text{cut,B} = 1.36$ for benchmark model B compared to the exponential overshooting grid and MF$_\text{cut,D} = 1.35$ for benchmark model D compared to the step overshooting grid.

Within the exponential diffusive overshooting grid, seven models have merit functions below MF$_\text{cut,A} = 1.41$ when compared against benchmark model A. Common for all seven models is that they have the same $M_\text{ini}$, $X_\text{ini}$ and $f_\text{ov}$ as the benchmark and $X_\text{c} = 0.5 \pm 0.003$. In other words, we cannot distinguish between models with exponential overshooting which differ in $X_\text{c}$ within 0.003 using g-modes. When compared to the more evolved benchmark model B, the number of models with merit functions below MF$_\text{cut,B} = 1.36$ increases to 51. For a given combination of $M_\text{ini}$ and $X_\text{ini}$ of these equally well matching models, we find that if $f_\text{ov}$ increases by 0.001 then $X_\text{c}$ simultaneuously decreases by $\sim 0.002$. Furthermore, $X_\text{ini}$ varies over the entire grid range whereas $M_\text{ini}$ is restricted to $3.15-3.3 \ \text{M}_\odot$ and $f_\text{ov}$ to $0.013-0.016$. In other words, the 51 models have initial stellar masses centered around the benchmark model B and $f_\text{ov}$ values skewed towards slightly lower values. 

In comparison, a lot more models are able to equally well match benchmark model C and D, for which a step overshoot description is used, within the step overshooting grid. For benchmark model C, 33 models return merit functions below the $\text{MF}_\text{cut,C} = 1.42$. Within these models we find that for a given combination of $M_\text{ini}$, $X_\text{ini}$ and $\alpha_\text{ov}$, $X_\text{c}$ varies with $\pm 0.002$. $X_\text{ini}$ varies over the entire grid range and slightly higher $M_\text{ini}$ ($3.2-3.35 \ \text{M}_\odot$) is generally favoured. For the overshooting parameter $\alpha_\text{ov}$, a similar or lower value is returned ($\alpha_\text{ov} = 0.10-0.16$). For the more evolved case of benchmark model D, 104 models fall below the cut-off MF$_\text{cut,D} = 1.35$. Similarly to the case of the exponential overshooting grid, for a given combination of $M_\text{ini}$ and $X_\text{ini}$ within these models when $\alpha_\text{ov}$ is increased by 0.01 then $X_\text{c}$ decreases by $\sim 0.002$. $X_\text{ini}$ varies over the entire grid range, whereas $M_\text{ini}$ is always equal to or higher than the initial mass of the benchmark model and $\alpha_\text{ov}$ is generally lower ($\alpha_\text{ov} = 0.10-0.16$).


\end{document}